\documentclass[a4paper,fleqn,usenatbib,useAMS]{mpazh}
\usepackage{graphicx}	% Including figure files
\usepackage{amsmath}	% Advanced maths commands
\usepackage{amssymb}	% Extra maths symbols
\usepackage{multicol}        % Multi-column entries in tables
\usepackage{bm}		% Bold maths symbols, including upright Greek

\usepackage{color}
\usepackage{pdflscape}
\usepackage[T1]{fontenc}
\usepackage{ae,aecompl}
\usepackage{newtxtext,newtxmath}
\usepackage{lscape}
\usepackage{mathtext}
\usepackage{float}
\usepackage{textcomp}
\usepackage{rotating}
\usepackage{dcolumn}
\usepackage{tabularx}
\usepackage{pdflscape}
\usepackage{url}
\usepackage{textcomp}

\renewcommand{\theenumi}{\roman{enumi}}%

\def\journal#1#2#3#4{{#1 {\bf #2}, #3 (#4)}}
\def\aj{Astron. J.}
\def\apj{Astrophys. J.}
\def\apjl{Astrophys. J. Lett.}
\def\apjs{Astrophys. J. Suppl. Ser.}
\def\apss{Astrophys. Sp. Sci.}
\def\aap{Astron Astrophys.}

\def\araa{ARAA}
\def\mnras{MNRAS}

\def\astl{Astron. Lett.}

\def\beq#1{\begin{equation}\label{#1}}
\def\eeq{\end{equation}}
\def\beqa#1{\begin{eqnarray}\label{#1}}
\def\eeqa{\end{eqnarray}}

\newcommand{\Msun}{M_\odot}
\newcommand{\Mbh}{M_{\rm BH}}
\newcommand{\Mb}{M_{\rm b}}
\newcommand{\Mstar}{M_\ast}

\newcommand{\ledd}{\lambda_{\rm Edd}}
\newcommand{\Lx}{L_{\rm X}}
\newcommand{\Lk}{L_K}
\newcommand{\Lksun}{L_{K,\odot}}
\newcommand{\nx}{n_{\rm X}}
\newcommand{\nk}{n_K}
\newcommand{\Lkagn}{L_{K,{\rm AGN}}}
\newcommand{\Lkg}{L_{K,{\rm g}}}
\newcommand{\Lkb}{L_{K,{\rm b}}}
\newcommand{\Pb}{P_{\rm b}}
\newcommand{\Le}{L_{\rm Edd}}
\newcommand{\Lbol}{L_{\rm bol}}

\newcommand{\lgl}{\log{\lambda_{\rm Edd}}}
\newcommand{\lglz}{\log{\lambda^*_{\rm Edd}}}

\title[SMBH Growth Rate and Its Mass Dependence]{The Growth Rate of Supermassive Black Holes and Its Dependence on the Stellar Mass of Galaxies at the Present Epoch }
 \author[Prokhorenko~et~al.]{Sergey~A.~Prokhorenko$^{2,1}$\thanks{E-mail: \href{mailto:sprokhorenko@iki.rssi.ru}{sprokhorenko@iki.rssi.ru}}, Sergey~Yu.~Sazonov$^{1}$
 \\
$^{1}$ Space Research Institute of the Russian Academy of Sciences (IKI), 84/32 Profsoyuznaya Str, Moscow, Russia, 117997 \\
$^{2}$ National Research University «Higher School of Economics», Pokrovsky Bulvar 11, 101990 Moscow, Russia}
\date{Accepted XXX. Received YYY; in original form ZZZ}

\pubyear{2021}

\begin{document}

\label{firstpage}
\pagerange{\pageref{firstpage}--\pageref{lastpage}}
\maketitle

\begin{abstract}
We study the distribution of accretion rates onto supermassive black holes in AGNs of the local Universe ($z<0.15$) based on near-infrared and hard X-ray surveys (2MASS and Swift/BAT). Using sufficiently accurate black hole mass estimates, we reliably estimated the Eddington ratio, $\ledd$, for approximately half of the objects in the AGN sample; for the remaining ones we used a rougher estimate based on the correlation of $\Mbh$ with the galaxy stellar mass $\Mstar$. We found that for a wide range of galaxy masses, $9.28<\log(\Mstar/\Msun)<12.28$, including the most massive galaxies in the local Universe, the distribution $f(\ledd)$ above $\lgl=-3$ can be described by a power law with $\Mstar$-independent parameters, declining with a characteristic slope $\approx 0.7$ up to the Eddington limit ($\lgl\sim 0$), where there is evidence for a break. In addition, there is evidence that at $\lgl<-3$ the dependence $f(\ledd)$ has a lower slope or flattens out. The mean characteristic growth time of supermassive black holes at the present epoch turns out to depend weakly on the galaxy stellar mass and to exceed the lifetime of the Universe, but by no more than one order of magnitude. The mean duty cycle of supermassive black holes (the fraction of objects with $\ledd>0.01$) in the local Universe also depends weakly on $\Mstar$ and is 0.2--1\%. These results obtained for the present epoch confirm the trends found in previous studies for the earlier Universe and refine the parameters of the dependence $f(\ledd|\Mstar)$ at $z<0.15$. The revealed universal (weakly dependent on the galaxy stellar mass) pattern of the dependence $f(\ledd)$ probably stems from the fact that, at present, the episodes of mass accretion onto supermassive black holes are associated mainly with stochastic processes in galactic nuclei rather than with global galaxy evolution processes.
\end{abstract}
%--------------------------------------------------------------------------------
%	Body
%--------------------------------------------------------------------------------
\section{Introduction}
\label{s:intro}
%--------------------------------------------------------------------------------
The nucleus of almost every galaxy at the present
epoch is currently believed to host a supermassive
black hole (SMBH). A correlation between the black
hole (BH) mass $\Mbh$ and such characteristics of the
galaxy central stellar bulge as the bulge mass $\Mb$ and
the velocity dispersion $\sigma$ has been revealed (Magorrian et al. 1998; Tremaine et al. 2002; for a review, see
Kormendy and Ho 2013). This suggests that the star
formation in galaxies and the growth of their central
BHs could correlate closely with each other during
the evolution of the Universe. In particular, the feedback mechanisms associated with huge energy release during mass accretion onto SMBHs could play
a prominent role (see, e.g., Ciotti and Ostriker 2001;
Di Matteo et al. 2005; Murray et al. 2005; Sazonov et al. 2005; for a review, see also Fabian 2012; King
and Pounds 2015; Naab and Ostriker 2017), despite
the fact that the BH mass typically accounts for only
$10^{-3}$--$10^{-2}$ of the bulge mass and the region of its
gravitational influence extends only to the galactic
nucleus. However, it is also clear that this correlation is not simple, as suggested, in particular, by
the detection of SMBHs in galaxies with pseudo-bulges and without bulges altogether (Kormendy and Ho 2013).

It is impossible to get a fairly comprehensive idea
of the correlation between SMBH growth and galaxy
evolution using individual objects as an example because of the huge (up to the cosmological ones) time
scales on which these processes occur. Therefore, it
is necessary to resort to statistical studies of galaxies
and active galactic nuclei (AGNs). Such studies
have been carried out particularly actively in the last
two decades owing to the appearance of sufficiently
deep surveys in various wavelength ranges. It has
been realized that the main growth of SMBHs in
the Universe occurred at redshifts $z\sim 1$--3 (see, e.g.,
Ueda et al. 2014; Aird et al. 2015), approximately
at the same epoch when the stars in galaxies were
formed most actively (for a review, see Madau and
Dickinson 2014), while at a later time both these
processes slowed down considerably. At the same
time, starting from $z\sim 3$ and up until now the integrated mass accretion rate onto BHs in galactic
nuclei was approximately proportional to the total
star formation rate in the Universe (see, e.g., Merloni
and Heinz 2008; Shankar et al. 2009). However,
these trends have been reliably established only for
the Universe on average, while much remains unclear
in the evolution of galaxies of various types and the
growth of their central BHs.

Most of the key questions remain open even for the
present epoch. For example, why are some galaxies
active (in particular, the Seyfert ones), while others
not? To what extent is this associated with the processes occurring in the central region of the galaxy
and with the evolution of the galaxy as a whole? What
is the characteristic AGN duty cycle? A possible
way of searching for answers to these questions is
a statistical study of the occurrence of AGNs with
different luminosities (i.e., with different accretion
rates onto SMBHs) in galaxies of different types. In
this case, the stellar mass $\Mstar$ can be used as a key
characteristic of galaxies.

One of the first such studies was carried out by
Aird et al. (2012). A representative sample of $\sim 25\times 10^3$ galaxies, among which there were $\sim 200$ X-ray-selected AGNs at redshifts $0.2<z<1.0$, was produced in several small sky fields (with a total area
$\sim3\, deg^2$). For all of the AGNs the accretion rate onto
the SMBH was estimated (from the measured X-ray
luminosity), while for all of the galaxies (including the
AGNs) the BH mass was roughly estimated under
the assumption that  $\Mbh=0.002\,\Mstar$ (assuming the
bulge to dominate in the total galaxy stellar mass).
As a result, the distribution of SMBHs in Eddington
ratio $\ledd$ (the ratio of the accretion rate to the critical
one specified by the Eddington luminosity) was studied, and this dependence [$f(\ledd)$] turned out to have
a falling power-law shape with a slope of about -0.65,
which does not depend on $\Mstar$, and a normalization
that increases with $z$. Thus, evidence that the SMBH
growth is universal in galaxies of different masses and
that it dramatically slowed down in the Universe as
a whole between the epochs $z=1$ and $z=0.2$ was
found.

Bongiorno et al. (2012) obtained results similar to those in Aird et al. (2012), but for the earlier Universe, $0.3<z<2.5$, and pointed out that
there is no dependence of the $\ledd$ distribution not only on the mass, but also on the star formation
rate; furthermore, a break in the power-law dependence $f(\ledd)$ was detected at $\ledd\sim 1$. Subsequently, Bongiorno et al. (2016) refined the shape
of this break and its evolution with redshift, while
Georgakakis et al. (2017) detected a flattening or,
possibly, a break of the function $f(\ledd)$ at $\lgl\lesssim -3$ and pointed out some dependence of $f(\ledd)$ on
$\Mstar$, whereby low accretion rates ($\ledd$) dominate 
among massive galaxies. Finally, while studying the
Universe in a wide range of redshifts, $0.1<z<4$,
Aird et al. (2018), on the whole, confirmed and refined
these trends.

In all of the mentioned papers the conclusions
about the distribution of accretion rates onto SMBHs
were drawn from a “specific accretion rate”, i.e., the
ratio of the AGN X-ray luminosity to the total host
galaxy stellar mass. However, since the correlation
between $\Mbh$ and $\Mstar$ is weak, the specific accretion
rate is only a rough approximation of $\ledd$. Furthermore, because of the limited area of the X-ray surveys
(no more than  $\sim 10^3\; deg^2$) used in these studies, the
range of their redshift coverage begins from $z\sim 0.1$--0.2, i.e., the present-day Universe turned out to be
least studied.

Hence, it should be noted that the dependence
$f(\ledd)$ at the present epoch was studied in Kauffmann and Heckman (2009), where the selection of
AGNs, only type 2 ones, was based not on X-ray
surveys, but on the optical Sloan Digital Sky Survey
(SDSS). More specifically, AGNs were selected by
their emission in narrow emission lines, while their
bolometric luminosities (and, as a consequence,
accretion rates) were estimated from the flux in the
narrow [OIII]$\lambda 5007$ line. The initial conclusions of
this paper differed radically from those of the studies
listed above, which are based on the X-ray selection
of AGNs. Subsequently, however, Jones et al. (2016)
showed that strong selection effects arise in this
method, while after the correction for them the results
based on SDSS agree satisfactorily with the fact
that the dependence $f(\ledd)$ (at low redshifts) has
a power-law shape with an exponential cutoff at
$\ledd\sim 1$.

The goal of this paper is to study the $\ledd$ distribution of SMBHs and its dependence on $\Mstar$ at the
present epoch ($z<0.15$). To systematically consider
as many AGNs in the local Universe as possible,
we will rely on two all-sky surveys: (1) the 2MASS
near-infrared photometric survey to construct a large
sample of galaxies ($\sim 10^6$ objects) with known redshifts and (2) the hard X-ray survey with the BAT
instrument onboard the Neil Gehrels observatory to
construct a sample of AGNs ($\sim 650$ objects). Using the near-infrared survey allows the galaxy stellar
masses to be estimated fairly accurately, while using the hard X-ray survey minimizes the influence of
selection effects when selecting AGNs. Apart from
rough SMBH mass estimates based on the correlation with the stellar bulge mass, we also use more
accurate $\Mbh$ estimates available approximately for
half of our AGN sample. As far as we know, such
a study is carried out for the first time for the local 
Universe.

%--------------------------------------------------------------------------------
\section{SELECTION OF OBJECTS FOR THE STUDY}
\label{s:data}
%--------------------------------------------------------------------------------
\begin{table*}
\begin{center}	

	\caption{Distribution of AGNs in types and sources of information about the redshift/distance}\label{typeagn} 
	
	\begin{tabular}{l|c|c|c|c|c} \hline\hline
		{Type}&  Total &Present in & Spectroscopic & Photometric  & Accurate \\ &
		 & 2MPZ+2MRS & $z$&$z$& distances\\ \hline
		{Seyfert I}   &   246      & 205 & 242 & 4      &  4 \\
		{Seyfert II}  &   353      & 339 & 343 & 10     & 41 \\
		{LINER }      &    5       &  5  & 5   & 0      &  1 \\
		{Unknown AGN} &    49      & 48  & 29  &20      &  1\\
		{Total:}      &    653     & 597 & 619 & 34     &  47 \\
		
 \hline
	
\end{tabular}
\end{center}
\end{table*}

\begin{figure*}
\begin{center}
	\vspace{6mm}
	\includegraphics[width=17cm]{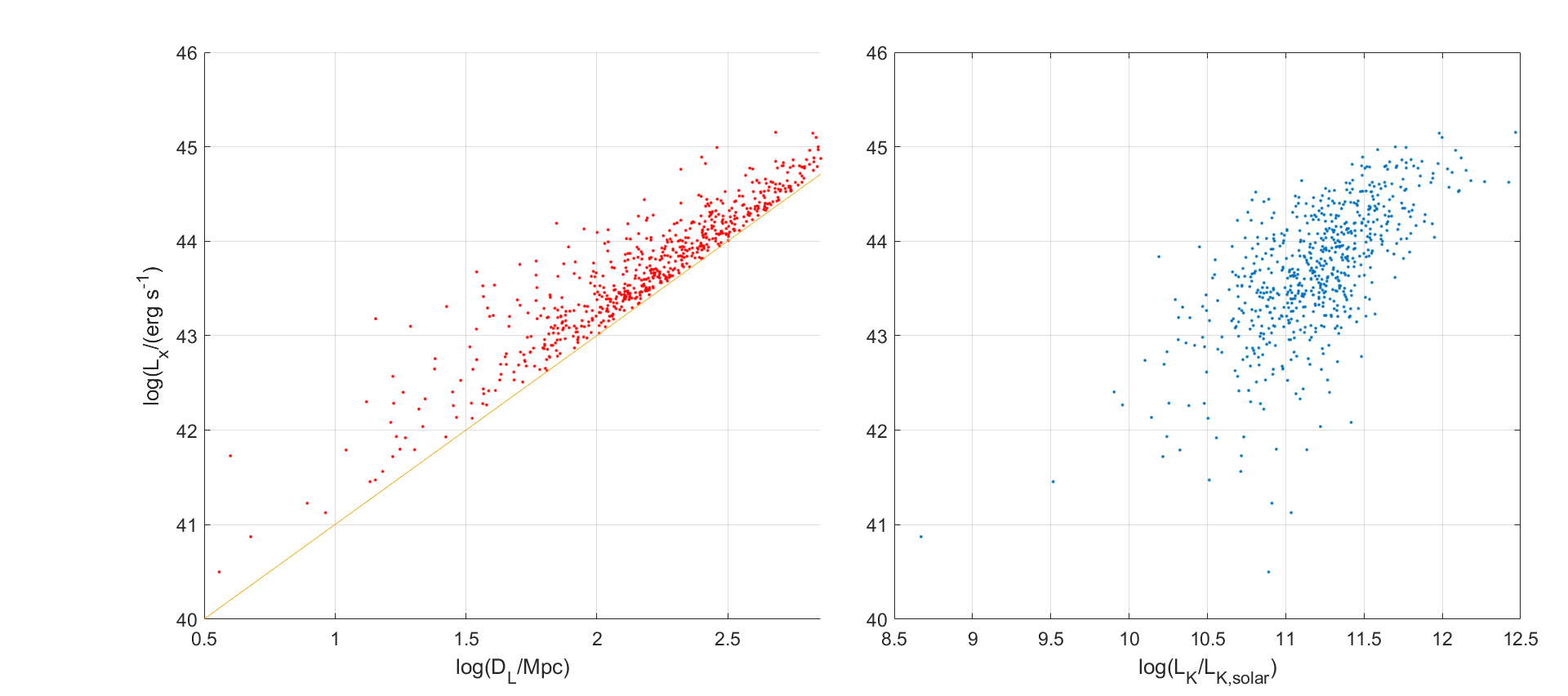}
	\caption{\rm \label{lxdl} {Left:} X-ray luminosity (14--195 keV) versus distance from the Swift/BAT AGN sample. The orange line indicates
the luminosity threshold corresponding to the minimum X-ray flux for an AGN to be included in the sample ($F_{\rm min}=8.4\times 10^{-12}$ erg s$^{-1}$ cm$^{-2}$).
{Right:} The distribution of AGNs in X-ray luminosity and $K$-band luminosity.}
\end{center}
\end{figure*}

\subsection{Galaxy Sample}
 
 The Two Micron All Sky Survey (2MASS) (Skrutskie et a. 2006) makes it possible to produce a large
homogeneous sample of galaxies in the relatively
nearby Universe and to estimate quite reliably their
stellar masses based on photometric measurements
in the $ K_s $ band ($ \lambda=2.159\,\mu m, \Delta \lambda=0.262\,\mu m$)\footnote{Below $ K_s $ will be abbreviated to $K$, despite the fact that it is
customary to consider a different band close in wavelength
(centered at $ \lambda=2.2\,
\mu m$) to be the K band}. To 
measure the galaxy luminosities in this band, we used
the 2MRS (Huchra et al. 2012) and 2MPZ (Bilicki
et al. 2013) catalogs of galaxy redshifts from 2MASS.
Spectroscopic redshift measurements are presented
in the 2MRS catalog; in the 2MPZ catalog there
are no spectroscopic redshift estimates for most of
the objects, but there are photometric $z$ estimates
found with machine learning algorithms. Although
such estimates are characterized by significant errors,
they are well suited for our purposes, given the more
serious assumptions that we have to make in the
course of our study.

Having cross-correlated the 2MRS (43533 galaxies) and 2MPZ (934175 galaxies) catalogs, we ascertained that the 2MPZ catalog lacks 4454 galaxies that are present in the 2MRS catalog. Therefore, to increase the statistical completeness of the
galaxy sample, it is necessary to use the combined
2MRS/2MPZ catalog. We excluded the sky region
near the Galactic plane ($ |b|<10^{\circ}$) from consideration, because in this zone the 2MRS and 2MPZ catalogs of galaxies are characterized by an insufficient
completeness. In addition, we imposed a constraint
in redshift, $z<0.15$. This was done again because
of the need to provide a high statistical completeness
of the sample and to be able to neglect the influence
of the cosmological evolution of the galaxy population
on the results of our study.

The above constraints ($|b|>10^{\circ}$ and $ z<0.15$)
are satisfied by 793289 galaxies from the combined
2MRS/2MPZ catalog.

\subsection{AGN Sample}

To produce a sample of AGNs, we used the
currently deepest available and homogeneous, over
most of the sky, catalog of hard X-ray (14--195 kev)
sources detected by the BAT instrument onboard
the Swift observatory (Swift/BAT). From the latest published version of the catalog based on the
first 105 months of Swift/BAT observations (Oh
et al. 2018) we selected AGNs with fluxes above $ 8.4\times 10^{-12}$ erg s$^{-1}$ cm$^{-2}$  (14--195~keV). At such a depth
90\% of the sky was covered (all sky was covered with
a sensitivity better than $ 9.3\times 10^{-12}$ erg s$^{-1}$ cm$^{-2}$). Similarly to the galaxy sample, we imposed the constraints  $|b|>10^{\circ}$ and $ z<0.15$ on the AGN sample as 
well. This choice additionally stems from the fact that
a significant fraction of the Swift/BAT X-ray sources
at low Galactic latitudes remain unidentified.

From the Swift/BAT catalog we took only objects with the Seyfert I (Sy 1.0-1.8), Seyfert II (Sy
1.9-2.0), LINER, and Unknown AGNs (i.e., Seyfert
galaxies and AGNs of an uncertain optical type) types
and excluded blazars (Beamed AGNs), because this
is a peculiar class of AGNs with highly collimated
emission toward the observer, which requires a separate consideration.

Thus, we selected 653 AGNs.

\subsection{Overlap between the Galaxy and AGN Samples}

\begin{figure*}
	\hspace{-2cm}
	\includegraphics[width=19cm]{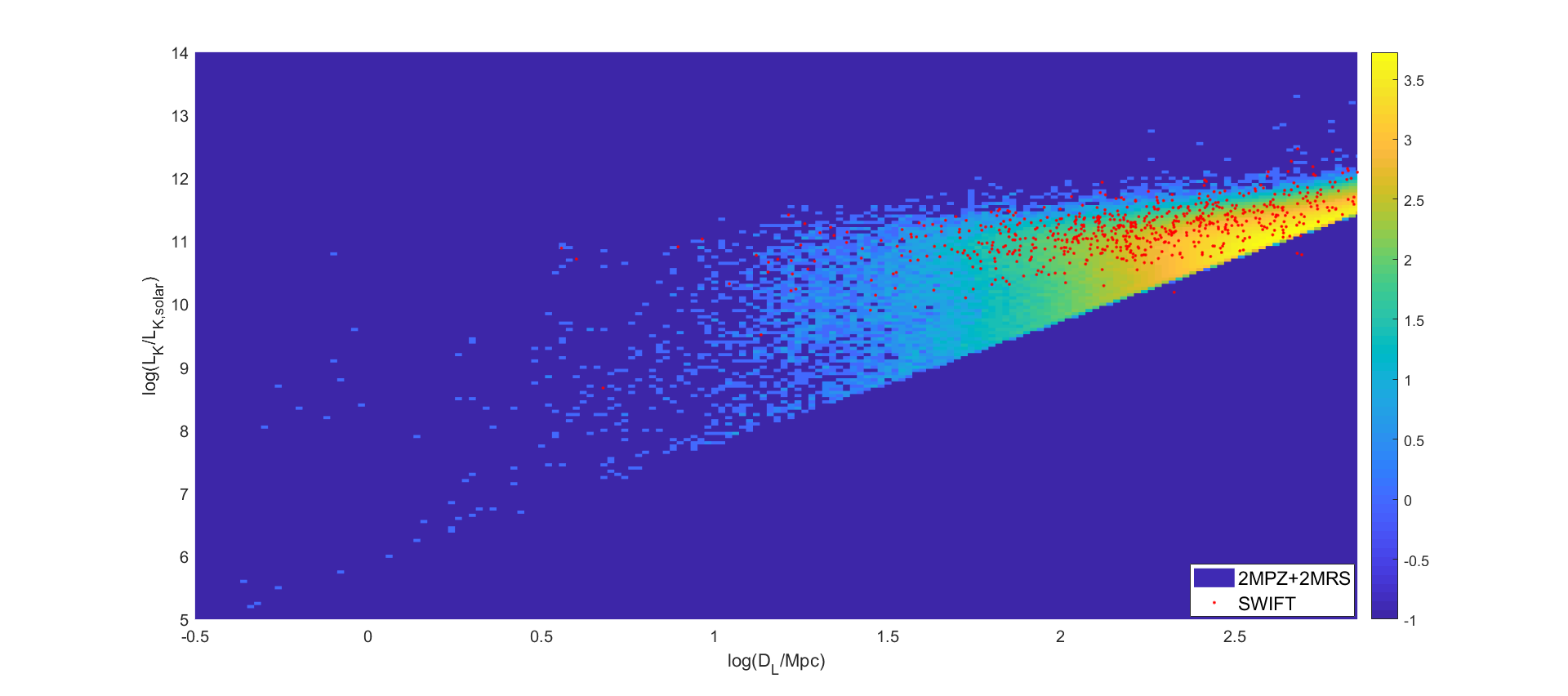}
	
	\caption{\rm \label{lkdl} $K$-band luminosity of the galaxies from the 2MRS/2MPZ sample (in solar luminosities in the same band) versus
distance. The field was divided into intervals with a width of 0.02 order of magnitude along the distance axis and 0.05 order of
magnitude along the luminosity axis. The color indicates the common logarithm of the number of galaxies that fell into a given
region of the diagram (in accordance with the color scale on the right). The value of the logarithm $-1$ means that no galaxy fell into a given region. The red dots mark the AGNs from the Swift/BAT sample.}
\end{figure*}

Having cross-correlated the Swfit/BAT AGN
sample and the 2MRS/2MPZ galaxy sample, we
found 597 matches (the controversial cases were
resolved manually using the NASA/IPAC Extragalactic Database (NED)). For 43 of these objects we
refined the $K$-band fluxes using the 2MASS Large
Galaxy Atlas (Jarrett et al. 2003), because it contains
more accurate measurements of the IR fluxes for
nearby galaxies.

For the remaining 56 AGNs (which are absent
in the 2MRS and 2MPZ catalogs) the $K$-band flux
was taken from the 2MASS Extended Catalog (Jarrett 2004) — 14 objects, the 2MASS All-Sky Catalog of Point Sources (Cutri et al. 2003) "— 41 objects, and the 2MASS Large Galaxy Atlas — 1 object
(a galaxy from the constellation Circinus).

\subsection{Distances to Objects}

The galaxy and AGN redshifts were taken from the
2MRS and 2MPZ catalogs using (if available) the
spectroscopic measurements and giving preference to
the 2MPZ catalog. For those AGNs that are absent
in the 2MRS and 2MPZ catalogs we used the redshifts given in the original catalog of the 105-month Swift/BAT survey. Note that the choice of a redshift
source is of no fundamental importance for the problem under consideration, because the spectroscopic
redshifts given in the Swift/BAT and 2MPZ catalogs
differ, on average, by 1.5\% and by more than 10\% only
in five cases (in these cases we took the estimates
from NED). The difference between the Swift/BAT
and 2MRS redshifts for the AGN sample is, on average, 1\% and does not exceed 7\% (in all but one case,
it is less than 5\%). For the AGN sample we took a 
total of 399 spectroscopic and 34 photometric $z$ from
the 2MPZ catalog and used the spectroscopic $z$ from
the 105-month Swift/BAT catalog for the remaining
220 objects.

For most of the galaxies and AGNs the photometric distances and, as a consequence, the luminosities
were calculated from the redshifts. In this case, we
used the cosmological model with $ \Omega_0=0.3 $ and $ h_0=70$~km~s$^{-1}$~Mpc$^{-1} $. For 47 nearest AGNs $z\lesssim 0.01$ we used more accurate distance estimates from the
Cosmicflows3 database (Tully et al. 2016). Similarly,
after the correlation of the 2MRS and 2MPZ catalogs
with Cosmicflows3, we corrected 8625 of the 42 533
and 7241 of the 934 175 distances to the galaxies in
these catalogs, respectively.

Thus, for all 653 AGNs in the sample we found
their counterparts in 2MASS, the corresponding $K$-band fluxes, and the distances. Information about the
optical types and redshifts/distances of the AGNs in
the sample is summarized in Table \ref{typeagn}.

\section{PROPERTIES OF THE GALAXY AND AGN
SAMPLES}

Figure \ref{lxdl}(left) shows the distribution of AGNs from the
sample under study in X-ray luminosity $\Lx$  (in the
14--195 keV energy band) and distance. Figure \ref{lxdl}(right)
presents the distribution of these AGNs in X-ray
luminosity and $K$-band luminosity $\Lk$. Here and 
below, the $K$-band luminosities of the objects are
expressed in solar luminosities in this band using the corresponding absolute magnitude of the Sun $ K_{\odot}=3.27 $ (Willmer 2018). As we see, the X-ray luminosities of the AGNs vary in a wide range, from $\Lx\sim 10^{41}$
to $\sim 10^{45}$~erg~s$^{-1}$, while their IR luminosities span a
range of more than two orders of magnitude.

Figure \ref{lkdl} shows the distribution of the sample of
galaxies from the combined 2MRS/2MPZ catalog in
$K$-band luminosity and distance; the AGNs from the
Swift/BAT sample are marked separately. As we see,
the selected AGNs are, on average, slightly brighter
in the infrared relative to the sample of all galaxies.

\subsection{Allowance for the AGN Contribution to the IR Luminosity of Galaxies}

\begin{figure*}
	\hspace{-1cm}
	\includegraphics[width=19cm]{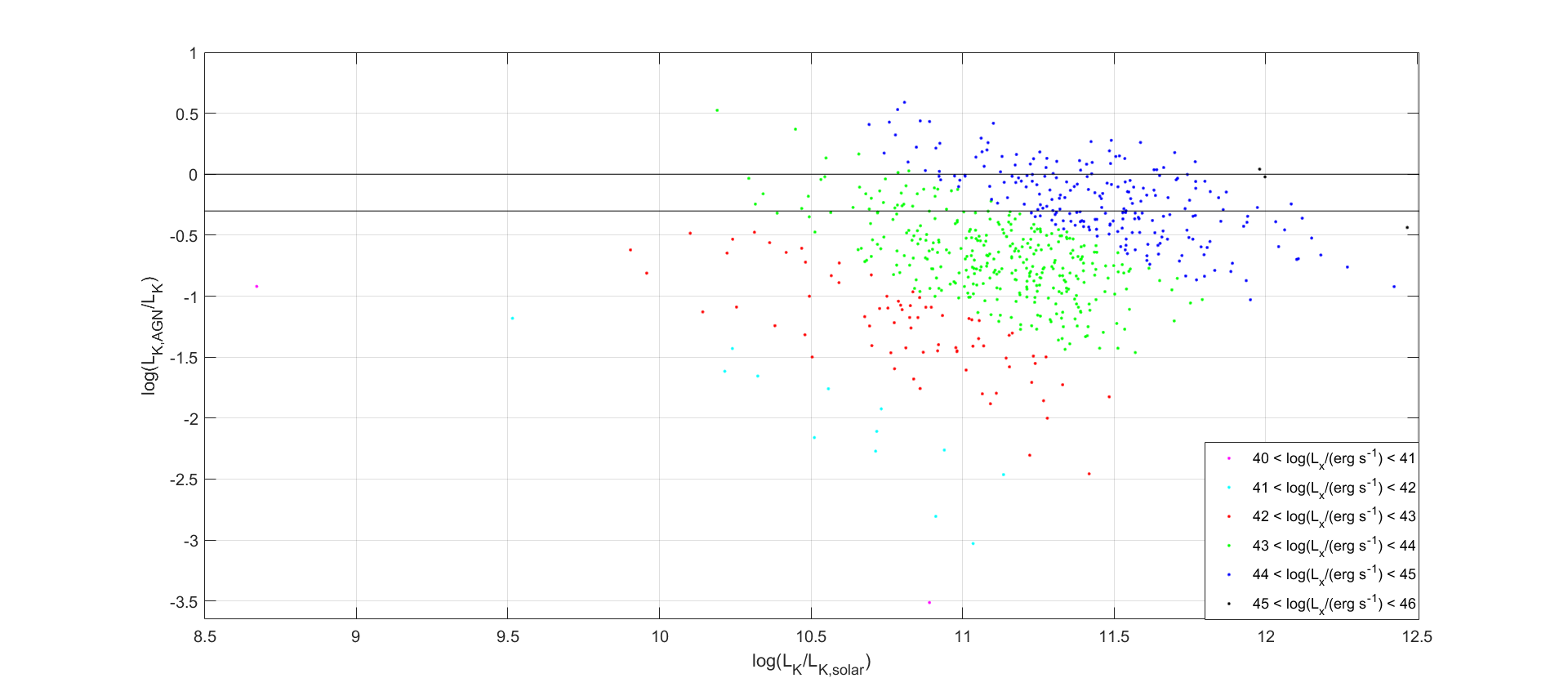}
	
	\caption{\rm \label{lkagn} Ratio of the expected AGN contribution to the total galaxy luminosity in the $K$-band for the AGN sample (as a function
of the luminosity). Different colors indicate objects with different hard X-ray luminosities. The two horizontal lines correspond
to the 50 and 100\% levels. Exceeding the 100\% level corresponds to an overestimation of the AGN contribution.}
\end{figure*}

\begin{figure*}

	\hspace{-2cm}
	\includegraphics[width=19cm]{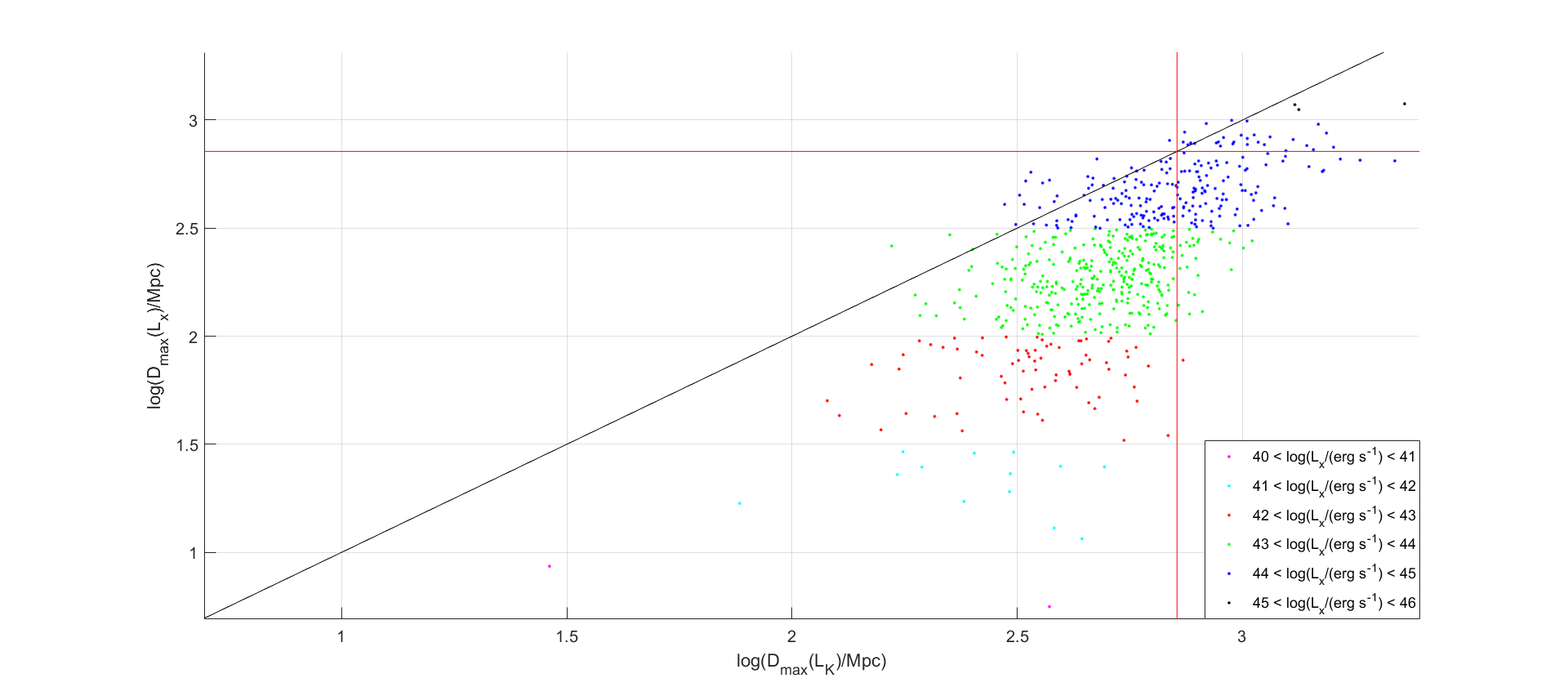}
	
	\caption{\rm \label{dmaxdmax} . Maximum viewable distances in the Swift/BAT hard X-ray survey and the 2MASS $K$-band survey for AGNs of different
luminosities (indicated by different colors) from the sample under study. The red lines mark the distances corresponding to $z=0.15$.}
\end{figure*}

Apart from the stellar population, the nucleus of
an active galaxy can contribute noticeably to its $K$-
band luminosity due to the absorption of part of the
emission from the SMBH accretion disk by the surrounding dust and gas and its reradiation in the infrared. This contribution, $\Lkagn$ , can be estimated
from the hard X-ray AGN luminosity using the correlation between these quantities. For the local AGN
population this relationship was derived by Sazonov
et al. (2012) based on data from the INTEGRAL sky
survey in the 17--60 keV energy band and Spitzer
Space Telescope observations and was converted to
the $K$ band by Khorunzhev et al. (2012). Adapting
Eq. (4) from the latter paper for the Swift/BAT energy
band (14–195 keV), we obtain
\begin{equation}
 \label{hor}
  \Lkagn \approx 0.05 \Lx.
\end{equation}
The conversion was made by assuming the X-ray
AGN spectra to have a power-law shape with a slope
$\Gamma=1.8$ and an exponential cutoff at 200~keV (for a review, see Malizia et al. 2020). It should be emphasized
that although the linear dependence (\ref{hor}) must describe satisfactorily the local AGN population on average, a
significant scatter of $\Lkagn/\Lx$ from object to object
is expected (Sazonov et al. 2012).

If we estimate the contribution of the nucleus to
the IR luminosity of the AGNs from our sample in
the described way, then the luminosities of their stellar
populations can be estimated as
\begin{equation}
\label{eq:lkg}
\Lkg=\Lk-\Lkagn.
\end{equation}
Figure \ref{lkagn}  shows the ratio $\Lkagn/\Lk$ for the AGN
sample. For 169 objects the expected contribution of the nucleus to the total galaxy luminosity in the
$K$ band is more than 50\%, while for 57 objects it
exceeds 100\%, suggesting an overestimation of the
contribution of the nucleus for a number of objects.
The related systematic uncertainty will be taken into
account below when studying the $\ledd$ distribution of
SMBHs.

\subsection{Luminosity Functions}

\begin{table*}
	
	\vspace{6mm}
	\centering
	\caption{Parameters of the best fit to the hard X-ray AGN luminosity function by the model (\ref{eq:lfagn}) and the $K$-band galaxy
luminosity function by the model (\ref{eq:lfgal})}\label{parlf} 
	
	\vspace{5mm}
	\begin{tabular}{c|c|c|c|c} \hline\hline
		{Luminosity}&$\phi^{*},$& $\gamma_1$ or  $\alpha$  &$\gamma_2 $ &$ L^*, $\ \\ {function}
		& Mpc$^{-3}$ \, dex$^{-1} $&            &          & erg s$^{-1}$\\ \hline
		AGNs   &     $ 2.2\,\pm\,0.2 \;\times\;10^{-5}  $        &$ 0.64  \,\pm\,_{0.08}^{0.06} $ &     $2.28 \,\pm\,_{0.06}^{0.02}$       &     $ 5.623 \, \pm \,_{1.157 }^{1.456}\;\times\;10^{43}  $     \\
		Galaxies &$7.89\,\pm\,0.01 \;\times\;10^{-3} $& $-1.053\,\pm\,0.01$            &            & $ 6.79 \, \pm 0.02\;\times\;10^{42}  $          \\
		\hline
		%\multicolumn{8}{}{\footnotesize a) значение нормировки учитывает неполноту выборки}
	\end{tabular}
\end{table*}

Using the samples of objects described above, we
can construct the luminosity functions of galaxies
in the near infrared and of AGNs in hard X-rays at
the present epoch ($z<0.15$). The standard $1/V_{\rm max}$
method is suitable for such calculations, where $ V_{\rm max}=(4\pi/3)\cdot0.826 D_{\rm max}^3 $ is the maximum
volume of the Universe in which an object with a
specified luminosity $\Lx$ or $\Lk$ could be detected in
the corresponding survey (Swift/BAT or 2MASS).
Here, the coefficient $0.826$ is equal to the fraction
of the total area of the celestial sphere at $|b|>10^\circ$. We determined the maximum distance $D_{\rm max}$
based on the Swift/BAT and 2MASS detection thresholds or, more specifically, the minimum flux
of $8.4\times 10^{-12}$ erg s$^{-1}$ cm$^{-2}$ in the 14--195 keV
band when constructing the X-ray AGN luminosity
function and the maximum apparent magnitude
$K=13.9$ (Bilicki et al. 2013) when constructing
the IR galaxy luminosity function. For the objects
whose maximum viewable distance $D_{\rm max}(\Lk)$ or
$D_{\rm max}(\Lx)$ exceeds $D_{L}(z=0.15)$, $V_{\rm max} $  was taken
to be $(4\pi/3)\cdot0.826 D_{L}^3(z=0.15) $. We calculated 
the errors in the specified luminosity bins based on
Poisson statistics as $ \pm\sqrt{\sum_{i}(1/V_{max,i}^2)} $.

Figure \ref{dmaxdmax} shows the distribution of AGNs from the
sample under study in maximum viewable distances
in the Swift/BAT hard X-ray survey and the 2MASS
$K$-band survey. We will need this two-dimensional
distribution below when studying the $\ledd$ distribution of SMBHs.

Apart from the discrete representation of the AGN
and galaxy luminosity functions, we attempted to
describe them by simple analytical models using the
maximum likelihood method with the following likelihood function:

\begin{equation}
{\mathcal{L}}=-2\sum_{i}\ln\frac{n(L_{{\rm obs},i})\,V_{\rm max}(L_{{\rm obs},i})}{\int n(L_{\rm obs})\,V_{\rm max}(L_{\rm obs})d\log\,L_{\rm obs}},
\label{eq:like}
\end{equation}

where $ L_{\rm obs} $ is the measured luminosity of the object
in the corresponding survey. The summation over $ i $
is done over all of the objects from the corresponding sample, while the normalization of the likelihood function is determined by the total number of objects in the sample.

To describe the X-ray AGN luminosity function
$ \nx (\Lx) $, we used a double power law:

\begin{equation}
\nx \equiv \frac{dN}{dV d\log\Lx} =\frac{\phi^*}{\left(\Lx/L^*\right)^{\gamma_1}+\left(\Lx/L^*\right) ^{\gamma_2}}, 
\label{eq:lfagn}
\end{equation}

which is commonly applied in AGN studies, while
to describe the $K$-band galaxy luminosity function
$\nk (\Lk) $, we used a Schechter function:

\begin{equation}
\nk \equiv \frac{dN}{dV d\log\Lk} =\phi^*\left( \frac{\Lk}{L^*} \right)^{\alpha+1} \exp\left(-\frac{\Lk}{L^*}  \right).
\label{eq:lfgal}
\end{equation}

\subsubsection{Allowance for the AGN Sample Incompleteness}

The sample of AGNs from the Swift/BAT survey
being considered here is not statistically complete.
First, there is a shortage of objects in 10\% of the
sky, because only 90\% is covered with a sensitivity
better than $ 8.4\times 10^{-12}$ erg s$^{-1}$ cm$^{-2}$. Second, in 
the original Swift/BAT catalog there are 83 objects
of an unknown nature at latitudes $|b|>10^{\circ}$, and if we 
assume that half of them are AGNs, then our sample
will increase by another $\sim 10$\%. This suggests that
the sample under study is incomplete approximately
by $20\%\pm10\%$. Therefore, when constructing the
AGN luminosity function and in all of the succeeding
calculations, the AGN space density was multiplied
by the coefficient 1.2, while the corresponding errors
were increased by 10\%.

\subsubsection{X-ray AGN Luminosity Function}

\begin{figure*}
	\hspace{-2cm}
	\includegraphics[width=19cm]{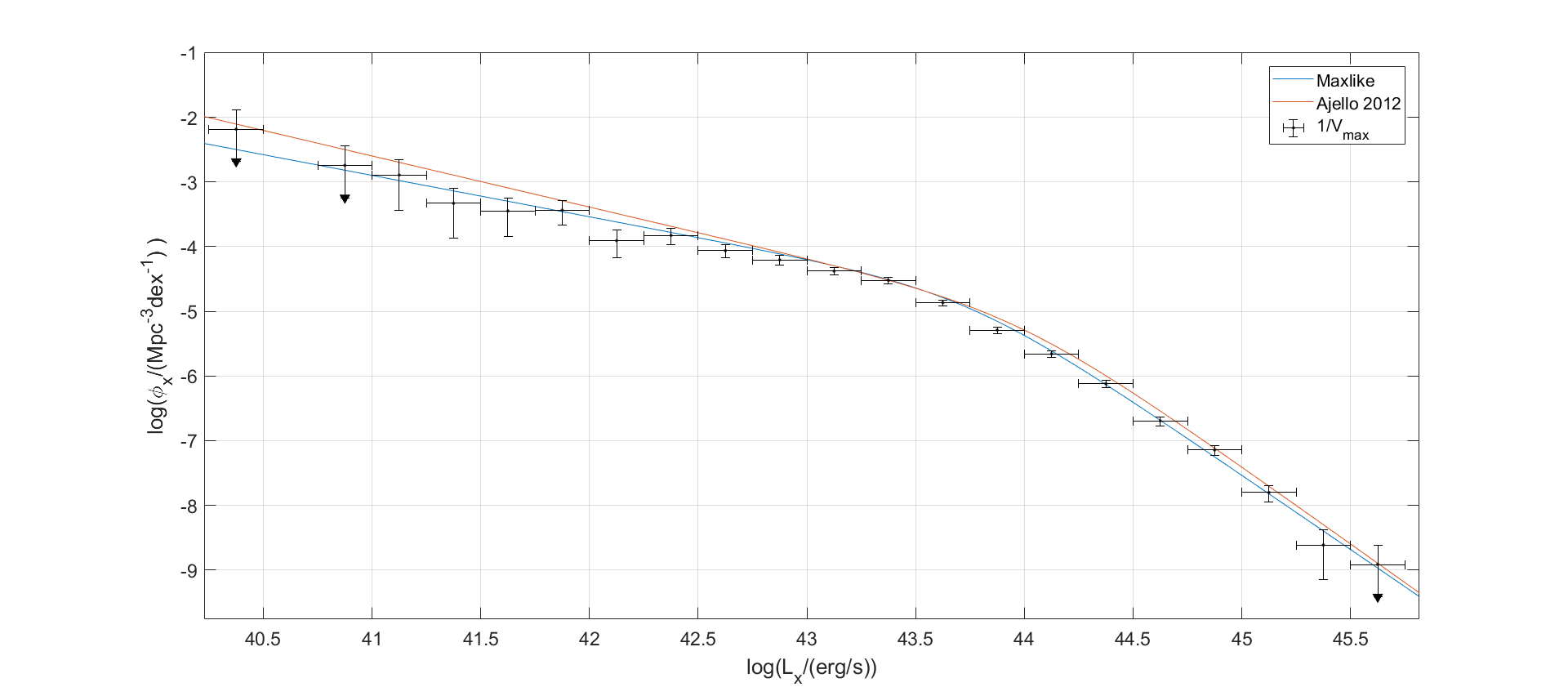}
	
	\caption{\rm \label{lx} The hard X-ray (14--195 keV) AGN luminosity function derived from the 105-month Swift/BAT survey data by the $1/V_{\rm max}$ method (black dots with error bars). The blue line indicates the best fit by a double power law calculated by the maximum likelihood method. For comparison, the red line indicates the model from \citep{Ajello} obtained from the 60-month Swift/BAT survey data.}
\end{figure*}

The hard X-ray
AGN luminosity function calculated by the $1/V_{\rm max}$
method and its best fit by the model (4) are shown in
Fig. (\ref{eq:lfagn}), while the best-fit parameters are presented in
Table \ref{parlf}.

As we see, the double power law model fits well the 105-month Swift/BAT survey data and
agrees satisfactorily with the result obtained previously by Ajello et al. (2012) based on a smaller sample
of AGNs from the 60-month Swift/BAT survey (for
this comparison, we converted the model parameters
given in Ajello et al. (2012) from the original 15--55 keV energy band to 14--195 keV using the model
of a power-law spectrum with a slope $\Gamma=1.8$ and an
exponential cutoff at 200 keV).

\subsubsection{$K$-band Galaxy Luminosity Function}

\begin{figure*}
	\hspace{-0cm}
	\includegraphics[width=17cm]{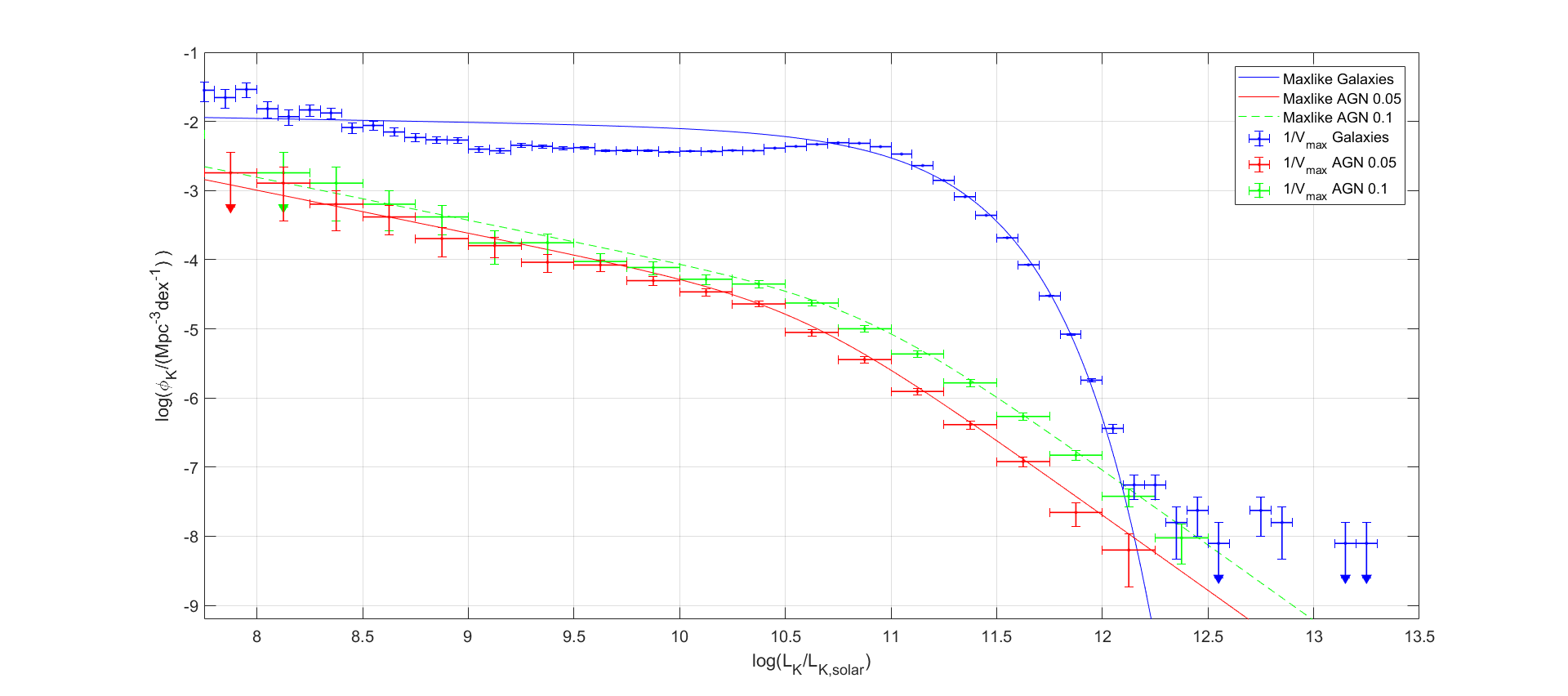}
	
	\caption{\rm \label{lk} The $K$-band galaxy luminosity function from the 2MASS data derived by the $1/V_{\rm max}$method (blue dots with error bars) and its best fit by the Schechter function calculated by the maximum likelihood method (blue line). Also shown are the $K$-band AGN luminosity functions derived by converting the corresponding hard X-ray luminosity function (Fig. \ref{lx}) using the
correlation (\ref{hor}) (red line) and a similar dependence, but with the coefficient 0.1 (green line).}
\end{figure*}

The $K$-band galaxy luminosity function calculated by the
$1/V_{max}$ method and its best fit by the model (\ref{eq:lfgal}) are
shown in Fig. \ref{lk}, while the best-fit parameters are
presented in Table \ref{parlf}..

Bonne et al. (2015) constructed the $K$-band
galaxy luminosity function based on a sample of
13\,325 nearby galaxies from the same 2MASS survey
as that in our paper. Although the model parameters
derived by us are close to those in Bonne et al. (2015)
($\phi^*=7.64^{+0.93}_{-0.83} \times 10^{-3}$~Mpc$^{-3}\,{\rm dex}^{-1}$, $\alpha=-1.17\pm0.08$, 
$\log(L^*/$erg s$^{-1})=42.974\pm0.024$), it can
be seen from Fig. \ref{lk} that the near-IR galaxy luminosity function is poorly described by the Schechter
function. For low luminosities this is because our
sample includes galaxies of both early (elliptical and
lenticular) and late (spiral) types, and it would be
more correct to describe the luminosity function
by the sum of two Schechter models with different
parameters (Bonne et al. 2015). In contrast, in the bright part of the luminosity function the observed
deviation of the estimates by the $1/V_{\rm max}$  method
from the analytical model probably stems from the
fact that a noticeable fraction of the very-high-luminosity galaxies in 2MASS may be active, and the
active nucleus (i.e., accretion onto the SMBH) can
contribute noticeably to their $K$-band luminosity.

The correlation (\ref{hor}) can be used to estimate the AGN contribution to the bright part of the $K$-band
galaxy luminosity function. For this purpose, we
need to simply shift the hard X-ray AGN luminosity
function (Fig. \ref{lx}) found above (from the Swift/BAT
data) by $|\log(0.05)|\,dex$ leftward along the luminosity
axis. The AGN contribution to the $K$-band galaxy
luminosity function estimated in this way is indicated
by the red line in Fig. \ref{lk}. We see that allowance
for this contribution makes it possible to explain the
observed bend of the galaxy luminosity function above
$\Lk\sim 10^{12.5}\Lksun$ only partially. This is most likely
due to a significant uncertainty in the fraction of the
bolometric AGN luminosity in the near infrared. For
example, if we repeat the estimation by increasing
the coefficient in Eq. (\ref{hor})  from 0.05 to 0.1, then it is
possible to describe the bend of the galaxy luminosity
function much better (see the green line in Fig. \ref{lk}).

As follows from Fig. \ref{lk}, up to $\Lk\sim 10^{12.5} \,\Lksun$
the AGN contribution to the $K$-band galaxy luminosity function may be deemed negligible, i.e., at
such luminosities the space density of ordinary (passive) galaxies is much higher than the AGN space
density. Therefore, since there are no objects with
$\Lk> 10^{12.5} \,\Lksun$ in the AGN sample under study
(from the Swift/BAT survey), in what follows we can
not only restrict ourselves to $\Lk< 10^{12.5} \,\Lksun$, but 
also assume that $\nk(\Lkg)\approx \nk(\Lk)$ for the galaxy
population (see Eq. (\ref{eq:lkg})).

Despite the fact that the simple Schechter model
describes poorly the $K$-band galaxy luminosity function, this in no way affects the results of our study,
because below we use only the nonparametric description of this function (the discrete estimates of the
galaxy space density obtained by the $ 1/V_{\rm max} $ method).

\section{EDDINGTON RATIO DISTRIBUTION OF
AGNs}

We now turn to the direct goal of our paper -- the
study of the Eddington ratio distribution of SMBHs
and its dependence of the galaxy mass.

\subsection{Mass-to-Light Ratio and Galaxy Stellar Mass
Function}

First of all we need to pass from the near-
IR luminosity to the galaxy stellar mass. The
mass-to-light ratio is known to depend on the
galaxy type. For example, for spiral galaxies McGaugh and Schombert (2014) obtained a typical
value of $\Mstar/\Lk=0.6\, \Msun/\Lksun$, while Martinsson
et al. (2013) provided the range of $\Mstar/\Lk=(0.31\pm0.07) \, \Msun/\Lksun$. In Bell et al. (2003) Fig. 20 presents
the B - R color–$\Mstar/\Lk$ correlation for galaxies
from 2MASS (i.e., galaxies of different types were
included). This correlation is weak: for most galaxies
$0.5\, \Msun/\Lksun < \Mstar/\Lk< 1.2\, \Msun/\Lksun$, while
the mean is $\approx 0.8\, \Msun/\Lksun$. Based on these observational data, below we will use a constant value,

\begin{equation}
\label{eq:mstar}
\frac{\Mstar}{\Lkg} = 0.6\frac{\Msun}{\Lksun},
\end{equation}

understanding that this ratio can actually vary by
$\sim 50$\% for galaxies from the sample under study.

Under this assumption we can pass from the
$K$-band galaxy luminosity function $\nk(\Lkg)\approx \nk(\Lk)$ to the galaxy stellar mass function $n_m(\Mstar)$
by shifting the argument of the function by $\log(0.6)$.

\subsection{Eddington Ratio}

It is convenient to describe the intensity of accretion onto SMBHs in terms of the ratio of the bolometric luminosity to the Eddington limit (Eddington
ratio):

\begin{equation}
\label{eq:ledd}
\ledd\equiv \frac{\Lbol}{\Le},
\end{equation}

where

\begin{equation}
 \label{Ledd}
  \Le = 1.3 \cdot 10^{38}\,\frac{\Mbh}{\Msun}
  \end{equation}
  
(erg~s$^{-1}$), and $\Mbh$ is the BH mass.

Based on a sample of AGNs (in the local Universe)
from the INTEGRAL hard X-ray survey, Sazonov
et al. (2012) showed that the luminosity of AGNs in
the 17--60 keV energy band is, on average, about
1/9 of their bolometric luminosity. Assuming (as
we have already done above) that the X-ray spectra
of AGNs are described by a power law with a slope
$\Gamma=1.8$ and an exponential cutoff at 200 keV, we
can write a similar relation to estimate the bolometric
AGN luminosity from the measured luminosity in the
Swift/BAT energy band (14--195 keV):

\begin{equation}
\label{eq:lbol}
\Lbol\approx 4.5\Lx.
\end{equation}

\subsection{Calculating the $\ledd$ Distribution of Galaxies}

Our goal is to determine the probability $ f(\ledd|\Mstar) $
that the SMBH in the nucleus of a galaxy with a
stellar mass $\Mstar$ accretes matter at a rate $\ledd$. This
quantity can be found as follows:

\begin{equation}
f(\ledd|\Mstar) = \frac{n_{ml}(\Mstar,\ledd)}{n_{m}(\Mstar)},
\label{eq:f}
\end{equation}

where

\begin{equation}
n_{ml}(\Mstar,\ledd)\equiv \frac{dN_{\rm gal}/dV}{d\log{\ledd}d\log{\Mstar}}
\end{equation}

is the galaxy space density per unit logarithmic interval of $\ledd$ and unit logarithmic interval of $\Mstar$ near
the specified $\ledd$ and $\Mstar$, respectively.

We calculated $n_{ml}(\Mstar,\ledd)$ by the $ 1/V_{\rm max} $ method,
where $ V_{\rm max} $ is the smaller of the volumes $V_{\rm max} (\Lkg)$
and $V_{\rm max} (\Lx)$ for a given AGN. The applicability of
this method and related uncertainties are discussed
below. It can be seen from Fig. \ref{dmaxdmax} that for the AGNs
of the sample under study the maximum viewable
volume $V_{\rm max}$ is determined mainly by the sensitivity
of the Swift/BAT X-ray survey rather than the IR
2MASS survey.

\subsubsection{Allowance for the Uncertainty in Estimating the
Stellar Masses of Galaxies with Active Nuclei}

The calculation of $n_{ml}(\Mstar,\ledd)$ consists of several steps:
\begin{enumerate}
    \item For each AGN (out of the 653) we estimate the
$K$-band galaxy stellar luminosity $ \Lkg $ from
Eq. (\ref{eq:lkg}) by taking into account the expected
contribution of the active nucleus to the galaxy
luminosity using Eq. (\ref{hor}) and then the stellar
mass $\Mstar$ from Eq. (\ref{eq:mstar}). In this step $\Lkg<0$ is
obtained for some objects and, naturally, these
are excluded from further consideration.
    \item For each of the remaining AGNs we calculate $ D_{\rm max} (\Lkg)$, i.e., the maximum distance to
which this object would be detected in 2MASS
if it had no active nucleus. If, as a result,
$ D_{\rm max} (\Lkg)$ turns out to be smaller than the
distance $D$ to the object, then this AGN is
excluded from further consideration.
    \item
     For each of the objects selected in the preceding steps we determine the maximum viewable volume $ V_{\rm max}(\Lkg,\Lx)=\min(V_{\rm max}(\Lx), V_{\rm max}(\Lkg)) $.
\end{enumerate}

Thus, we got three groups of AGNs:\\
(1) 536 objects with $ D<D_{max} (\Lkg)$ and $ \Lkg >0 $,\\
(2) 60 objects with $ D>D_{max} (\Lkg)$ and $ \Lkg >0 $,\\
(3) 57 objects with $ \Lkg \le 0$.

The presence of objects with $\Lkg \le 0$ suggests
that we overestimated the contribution of the active
nucleus to the $K$-band luminosity of these galaxies.
Previously (Fig. \ref{lkagn}) we have already noted that the
$\Lkagn$ estimates are characterized by a significant
uncertainty. To take into account the influence of
this factor on the results of our study, we calculated
$n_{ml}(\Mstar,\ledd)$ by three methods.

The first method is based on the algorithm described above. Specifically, we take into account the
contribution of the active nucleus to the IR galaxy
luminosity based on Eq. (\ref{hor}) and use the subsample
of 536 AGNs with $ D<D_{\rm max} (\Lkg)$ and $ \Lkg >0 $ in
our calculation. In this case, the AGN stellar masses,
on average, are somewhat underestimated.

We also performed an alternative calculation of
$n_{ml}(\Mstar,\ledd)$ by completely neglecting the contribution of the active nucleus when calculating the
galaxy masses, i.e., by assuming $\Lkg=\Lk$. In this
case, the total sample of 653 AGNs is used in our
calculation, but the galaxy stellar masses are overestimated.

Finally, we performed a calculation by the third
method, which is a modification of the first one.
Specifically, for all of the objects with $ \Lkg \le 0$ the
$K$-band galaxy luminosity was estimated as $ \Lkg^*=0.5\Lk$(in the remaining cases, as above, $\Lkg^*=\Lk-\Lkagn$). In this case, the final sample for
calculating $n_{ml}(\Mstar,\ledd)$ was 556 objects (for the
remaining 97 $ D >D_{\rm max} (\Lkg^*)$).

The third method may be deemed intermediate
between the first and second ones, which probably
give extreme estimates of $n_{ml}(\Mstar,\ledd)$.
The errors of $n_{ml}(\Mstar,\ledd)$ in each specified interval of $\Mstar$ and $\ledd$ were calculated as $\Delta_{\rm tot}=\sqrt{\Delta^2_{\rm stat}+\Delta^2_{\rm syst}} $. We estimated the statistical error
$\Delta_{\rm stat}$ as $\sqrt{\sum1/V_{\rm max}^2} $
and the systematic one (the
error of the method) $\Delta_{\rm syst}$ as the difference between
the extreme estimates obtained by the first and
second methods of estimating the AGN contribution
described above. The final error of $f(\ledd|\Mstar)$ was
calculated according to the definition (\ref{eq:f}), as $\Delta f\approx \Delta n_{ml}/n_m+\Delta n_m  \frac{n_{ml}}{n_m^2}$, i.e., the error of the galaxy
stellar mass function $n_m(\Mstar)$  was deemed negligible.

\subsubsection{Estimating the SMBH Masses}

\begin{table*}
	
	\vspace{6mm}
	\centering
	\caption{The number of AGNs with “accurate” SMBH mass estimates obtained by different methods }\label{typesmbh} 
	
	\vspace{5mm}
	\begin{tabular}{c|c|c|c|c} \hline\hline
		 Reverberation mapping & From H$\alpha$ & From H$\beta$ & From velocity dispersion & Total  \\
		 \hline
		 39       &   179        & 149 & 164 & 332 \\

 \hline
		
	\end{tabular}
\end{table*}

\begin{figure*}
	\hspace{-0cm}
	\includegraphics[width=17cm]{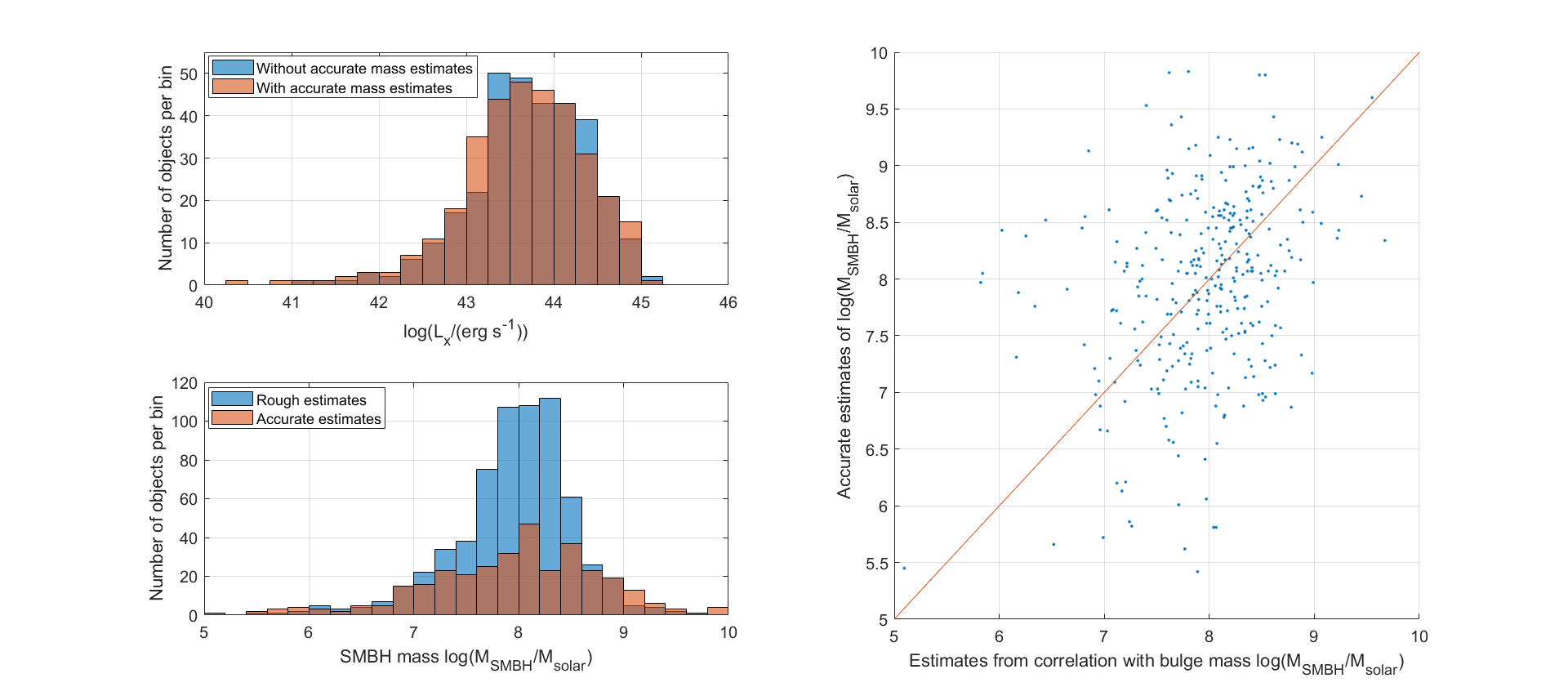}
	
	\caption{\rm \label{gistmbh} {Top left:} Hard X-ray luminosity ($ \Lx$) distribution of the AGNs for which only the SMBH mass estimate from the $K$-band galaxy luminosity (“rough estimate”) is available (321 of the 653 objects) and those for which more accurate SMBH mass
estimates are available (332 of the 653 objects). The bin width is 0.25 order of magnitude. {Bottom left:} The $\Mbh$ distribution
of rough SMBH mass estimates for all of the sample AGNs and more accurate estimates for the subsample of 332 objects.
The bin width is 0.2 order of magnitude 
{Right:} Comparison of the rough and accurate $\Mbh$ estimates for the subsample of 332 AGNs.}
\end{figure*}

Despite the fact
that the sample of AGNs from the Swift/BAT survey
used in this paper consists of relatively nearby objects
($z<0.15$) and has been studied reasonably well, the
estimates of the BH masses in these objects are characterized by a significant uncertainty. To estimate the
influence of this factor on the results of our study, we
used two types of $\Mbh$ estimates for the AGN sample.

First, for all AGNs $\Mbh$ were estimated from the
known correlation of the SMBH mass with the $K$-band stellar bulge luminosity $\Lkb$ (i.e., basically,
from the correlation with the bulge mass (Kormendy
and Ho 2013)):

\begin{equation}
 \label{Mbh}
   \frac{\Mbh}{10^9\Msun}=0.544\left(\frac{\Lkb}{10^{11}\Lksun}\right)^{1.22}.
\end{equation}

The bulge luminosity was estimated as $\Lkb=\Pb\Lkg$, where $\Pb$ is the fraction of the bulge in the
total stellar mass of the galaxy. The latter depends
on the morphological type of the galaxy. It should
also be kept in mind that relation (\ref{Mbh}) was derived for
galaxies with classical bulges, while its application
in the case of galaxies with pseudo-bulges can lead
to a significant error in estimating the BH masses
(Kormendy and Ho 2013). Because of the lack of
information about the morphology of the objects
under study, we fixed $\Pb$  at $0.25$ for all of the AGNs
from the Swift/BAT survey. Note that $\Pb=1$ for elliptical galaxies, $\Pb\approx 0.25$--$0.2$ for S0--Sb galaxies,
and $\Pb\approx 0.08$ for Sc galaxies (see, e.g., Laurikainen
et al. 2007, 2010; Graham and Worley 2008). Below
the estimates based on Eq. (\ref{Mbh}) are called “rough
estimates.”

Second, for AGNs we used the $\Mbh$ estimates
obtained by more accurate methods from Koss et al.(2017) and Marchesini et al. (2019) (see Table \ref{typesmbh}).
Such estimates (hereafter “accurate estimates”) exist
approximately for half of the AGN sample under study
or, more specifically, for 332 of the 653 objects. If
several different $\Mbh$ estimates were available, then
we preferred those obtained by the reverberation mapping method and, subsequently, in order of priority:
the estimates from the width and luminosity of the
broad H$\alpha$ emission line, from the width of the H$\beta$
emission line and the optical continuum luminosity,
and from the correlation of the SMBH mass with the
stellar velocity dispersion in galaxies.

Figure \ref{gistmbh} (top left) shows the hard X-ray luminosity distribution of the AGNs for which only rough
SMBH mass estimates are available and those for
which accurate estimates are available. As we see,
the fraction of AGNs with accurate estimates in the
sample under study is virtually independent of $\Lx$.
The same figure (bottom left) shows the distribution
of rough and accurate $\Mbh$ estimates. The second
distribution is seen to be wider (for example, judging
by the full width at half maximum of the distribution)
than the first one.

The existing uncertainty in the SMBH mass estimates for the AGN sample is best demonstrated by
the diagram presented on the right in Fig. \ref{gistmbh}, where the
rough and accurate $\Mbh$ estimates for the subsample
of 332 AGNs are compared directly. There is a large
scatter: the rms deviation between the logarithms of
the rough and accurate estimates is $0.86\,dex$.

To take into account the influence of the uncertainty in the SMBH mass estimates in AGNs on the results of our study, we calculated $n_{ml}(\Mstar,\ledd)$ by
three methods (in addition to the three different methods of allowance for the active nucleus described in
the previous subsection when calculating the stellar
mass).

The first method consists in using the rough $\Mbh$
estimates for the entire AGN sample (653 objects).
The second one consists in using only the accurate
estimates for the subsample of 332 objects. The
previously noted fact (Fig. \ref{gistmbh}) that the fraction of the
AGNs in the sample under study for which there are
accurate SMBH mass estimates is virtually independent of the X-ray luminosity, i.e., no noticeable
selection effect with regard to this characteristic is
observed, serves as a basis for the applicability of this
approach. In this case, however, the incompleteness
coefficient $\kappa=653/332$ is required to be introduced
when calculating $n_{ml}$:

\begin{equation}
\label{eq:acc}
n_{ml}(\Mstar,\ledd)=\kappa\; \sum_{i}\frac{1}{V_{\rm max,i}}.
\end{equation}

The summation in Eq. (\ref{eq:acc}) is over all of the AGNs in
a given interval in $\Mstar$ and $\lgl$. Accordingly, the
statistical errors are calculated as $ \pm\sqrt{\kappa\sum 1/V_{\rm max,i}^2} $.

The third method (”mixed estimates”) is based
on the accurate $\Mbh$ estimates for the subsample of
332 AGNs and the rough estimates for the remaining
321 sample objects.

As regards the applicability of the $1/V_{\rm max}$,
the volume in which all of the AGNs with specified
$\Lkg$ (or, identically, $\Lkg^*$) and $\ledd$ can be detected should be used as $V_{\rm max}$ in the first method of estimating the SMBH masses. Since the X-ray luminosity $\Lx$ is uniquely determined via $\Lkg$ and $\ledd$
in the rough SMBH estimate, this is equivalent to
the volume in which all of the AGNs with given $\Lkg$
and $\Lx$ can be detected. For the second method the
viewable volume for each AGN with an accurate $\Mbh$
may be deemed to be the same as that in the first
method, but (supposing that in this volume $\ledd$ for
the AGNs without accurate $\Mbh$ are distributed in
the same way as those for the objects with accurate
$\Mbh$ estimates, based on the top left histogram in
Fig. \ref{gistmbh}) then we should assume that it contains a factor of $\kappa$ more objects. The third method supposes to take the
viewable volume to be the same as that in the first
method. Here, there is no need to make a correction for incompleteness, because the incompleteness
of the objects with accurate estimates in a specific
volume is compensated for by the objects with rough
estimates.

\section{RESULTS}

\begin{table*}
	
	\vspace{6mm}
	\centering
	\caption{The best fits to the dependence $ f(\ledd|\Mstar) $ by a power law in $\ledd$ with $\Mstar$-independent parameters (Eq. (\ref{ffitl}))
for different SMBH mass estimates. The most probable values of the parameters and (in parentheses) the $1\sigma$ confidence
intervals are given}\label{bl} 
	
	\vspace{5mm}
	\begin{tabular}{l|c|c|c} \hline\hline
	{Best-fit parameter}& From rough estimates & From accurate estimates & From mixed estimates  \\
	\hline
	 Normalization $\log(A/$dex$^{-1})$ & $-5.07\;(-5.19,\; -4.95)$	&$-3.55\;(-3.66,\;-3.45) $  & $-3.79\;(-3.93,\;-3.68)$   \\\hline
    Slope $\gamma$ & $-1.3\;(-1.36,\;-1.24)$&$-0.71\;(-0.77,\;-0.65)$ &$-0.82\;(-0.91,\;-0.76)$ \\ \hline
    $\chi^2$ at minimum & 35.78  &28.1  & 15 \\ \hline
    Number of degrees &24  &14  & 18 \\ 
    of freedom $n-k$ &  & & \\ \hline
    $AIC_c$ &40.3  &33.02  & 19.7 \\ \hline		
    $BIC$ &42.29  &33.65 & 20.99 \\ \hline
	\end{tabular}
\end{table*}

\begin{table*}
	
	\vspace{6mm}
	\centering
	\caption{The best fits to the dependence $ f(\ledd|\Mstar) $ by a Schechter function in $\ledd$ with $\Mstar$-independent parameters (Eq. (\ref{ffitc})) for different SMBH mass estimates}\label{bsh} 
	
	\vspace{5mm}
	\begin{tabular}{l|c|c|c} \hline\hline
	{Best-fit parameters} & From rough estimates & From accurate estimates & From mixed estimates  \\ 
	\hline
	 Normalization $\log(A/$dex$^{-1})$ & $-2.13\;(-2.35,\; -1.96)$	&$-3.07\;(-3.35,\;-2.85) $  & $-3.72\;(-4.28,\;-3.35)$   \\\hline
    Slope $\gamma$ & $-0.56\;(-0.72,\;-0.39)$&$-0.44\;(-0.54,\;-0.33)$ &$-0.67\;(-0.79,\;-0.55)$ \\ \hline
    $\lglz$ & $-1.68\;(-1.77,\;-1.58)$&$0.16\;(-0.04,\;0.34)$ &$0.35\;(0.02,\;0.8)$ \\ \hline
    $\chi^2$ at minimum &17.3  &18.65  & 12.34 \\ \hline
    Number of degrees &23  &13  & 17 \\ 
    of freedom $n-k$ &  & & \\ \hline
    $AIC_c$ &24.39  &26.65  & 19.84\\ \hline		
    $BIC$ &27.07  &26.96 & 21.33 \\ \hline
	\end{tabular}
\end{table*}

\begin{table*}
	
	\vspace{6mm}
	\centering
	\caption{The best fits to the dependence $ f(\ledd|\Mstar) $ by a power law in $\Mstar$ and by a Schechter function in $\ledd$ (Eq. (\ref{ffitksh})), for different SMBH mass estimates }\label{bksh} 
	
	\vspace{5mm}
	\begin{tabular}{l|c|c|c} \hline\hline
	{Best-fit parameters} & From rough estimates & From accurate estimates & From mixed estimates  \\ 
     \hline
	 Normalization $\log(B/$dex$^{-1})$ & $-2.34\;(-2.82,\; -2.02)$	&$-3.53\;(-4.34,\;-2.36) $  & $-3.89\;(-5.36,\;-3.26)$   \\\hline
    Slope $\theta$	 &$-0.24\;(-0.38,\;-0.1)$ &$-0.33\;(-0.51,\;-0.08)$  &$-0.11\;(-0.26,\;0.05)$  \\\hline
    Slope $\gamma$ & $-0.7\;(-0.95,\;-0.46)$&$-0.6\;(-0.74,\;-0.45)$ &$-0.71\;(-0.88,\;-0.54)$ \\\hline
    $\lglz$ & $-1.61\;(-1.76,\;-0.37)$&$0.35\;(-0.2,\;1.04)$ &$0.43\;(-0.16,\;1.8)$ \\\hline
    $\chi^2$ at minimum &11.17  &15.15  & 11.37 \\ \hline
    Number of degrees &22  &12  & 16 \\ 
    of freedom $n-k$ &  & & \\ \hline
    $AIC_c$  &21.08  &26.78  & 22.04 \\ \hline		
    $BIC$ &24.2  &26.24 & 23.36 \\ \hline	
	\end{tabular}
\end{table*}

\begin{figure*}
	\hspace{-0cm}
	\includegraphics[width=17cm]{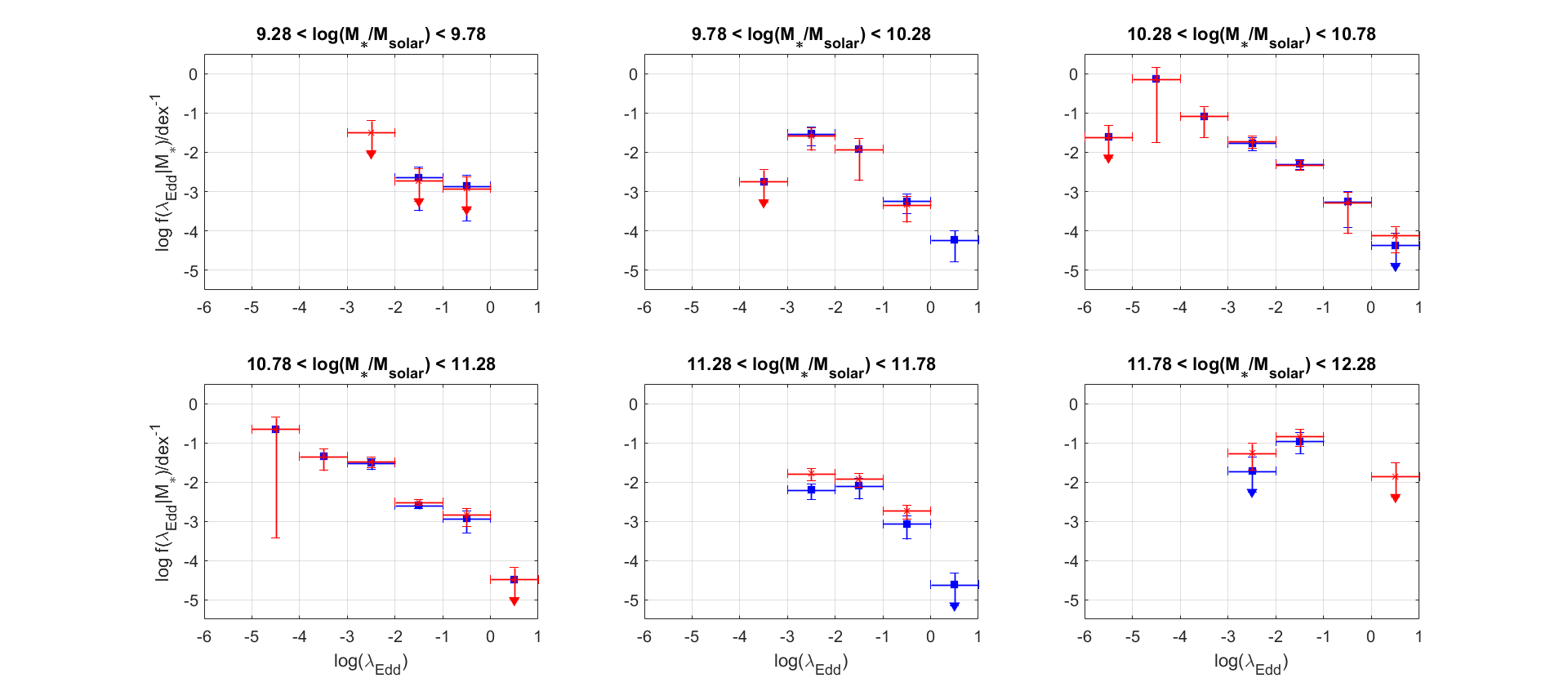}
	
	\caption{\rm \label{leddmbhlu} The dependence $f(\ledd)$ for different intervals of the galaxy stellar mass $\Mstar$, calculated by the $1/V_{\rm max}$. method. The results obtained by using the two extreme methods of allowance for the active nucleus in calculating  $\Mstar$ are compared: the first (based on $\Lkg$, blue dots) and the second (based on $\Lk$, red dots). In both cases, we used mixed (rough and accurate) SMBH mass estimates.}
\end{figure*}

Figure \ref{leddmbhlu} shows the calculated dependence
$ f(\ledd|\Mstar)$ in six galaxy stellar mass intervals of
width $0.5 dex$ from $\log(\Mstar/M_\odot)=9.28$ to $12.28$. This
range is determined by the boundaries of the $K$-band
luminosity range for the AGNs of the sample under
study ($8.5<\log(\Lk/\Lksun)<12.5$, see, e.g., Fig. \ref{lxdl})
and the $\Lkg$ -to-$\Mstar$ conversion factor in Eq. (\ref{eq:mstar}),
but we additionally limited the luminosity range from
below by $\log(\Lkg/\Lksun)=9.5$, because below this value
there is only one AGN in our sample. Different colors in the figure indicate the dependences derived by
using the two extreme methods of allowance for the
active nucleus in calculating the galaxy stellar mass:
the first (when $\Mstar$ is estimated from $\Lkg$) and the
second (when the total galaxy luminosity $\Lk$ is used
instead of $\Lkg$). In this case, to calculate $\ledd$, we
used the mixed $\Mbh$ estimates (third method).

In most of the $\log\Mstar$ intervals $ f(\ledd)$ clearly
shows a falling trend, with the uncertainty due to the
influence of the active nucleus on the $K$-band galaxy
luminosity having no strong influence on the shape of
the dependence.

\begin{figure*}
	\hspace{-0cm}
	\includegraphics[width=17cm]{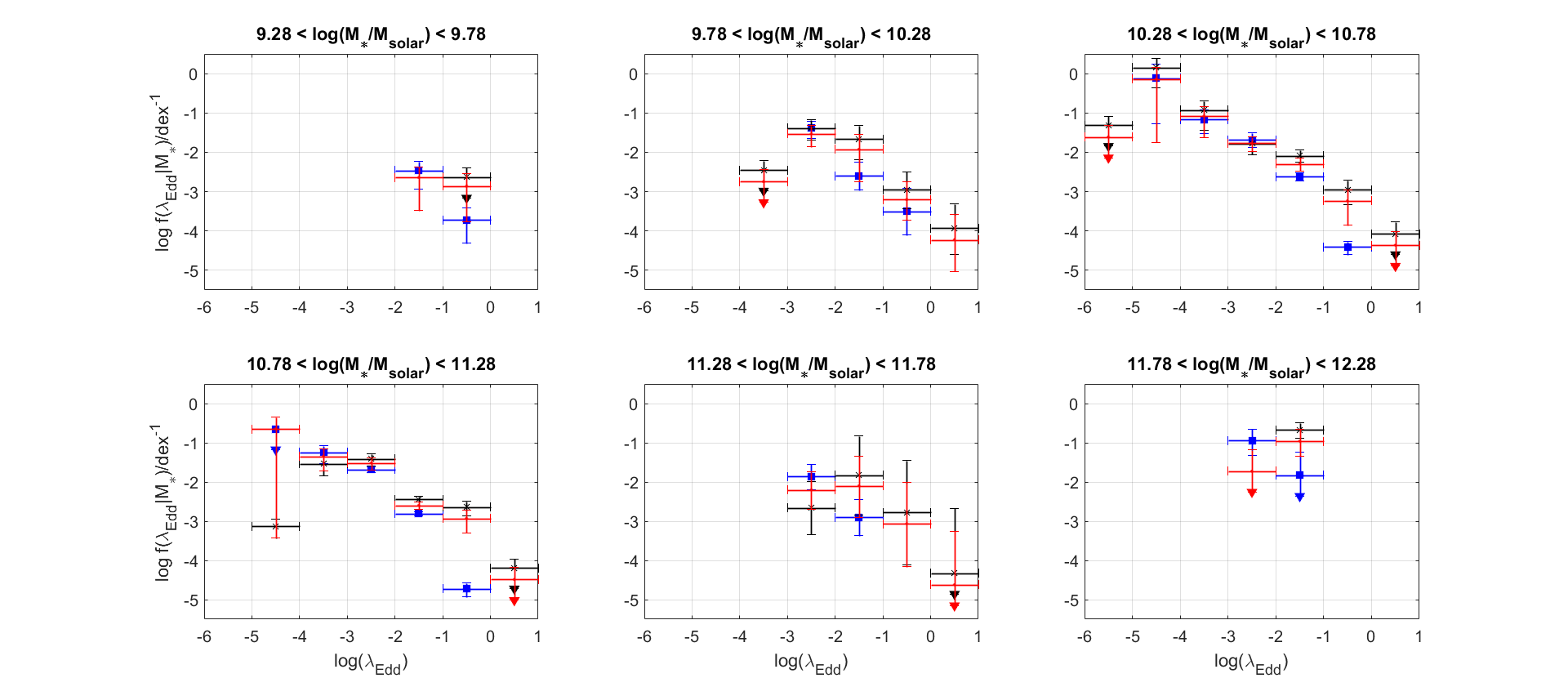}
	
	\caption{\rm \label{leddmbh} . Same as  (Fig. \ref{leddmbhlu}), but for different types of SMBH mass estimates: the rough (blue), accurate (black), and mixed (red) ones. The intermediate method of allowance for the active nucleus (using $\Lkg^*$, see the text) was used everywhere in calculating
the stellar mass.}
\end{figure*}

Figure \ref{leddmbh} shows the analogous dependences derived by the intermediate method of allowance for
the active nucleus when calculating the galaxy stellar
mass (using $\Lkg^*$) and three different methods of
estimating the SMBH masses or, more specifically,
based on the rough, accurate, and mixed $\Mbh$ estimates. We see a significant scatter of values due to
the uncertainty in the $\Mbh$ estimates for the AGN
sample under study.

Given that the dependence $ f(\ledd)$ falls off, we
attempted to describe it by a power law with a slope
and normalization independent of the stellar mass:

\begin{equation}
\label{ffitl}
f(\ledd|\Mstar)=A\ledd^\gamma.
\end{equation}

It was fitted using the $ \chi^2$ test based on the data points
estimated by the $ 1/V_{\rm max} $ method, given the corresponding errors. The calculation was performed by
the intermediate method of allowance for the contribution of the active nucleus to the IR galaxy luminosity and by taking into account the related statistical uncertainty (as was described in the Subsection “Allowance for the Uncertainty in Estimating the Stellar
Masses of Galaxies with Active Nuclei”).

\begin{figure*}
	\hspace{-2cm}
	\includegraphics[width=19cm]{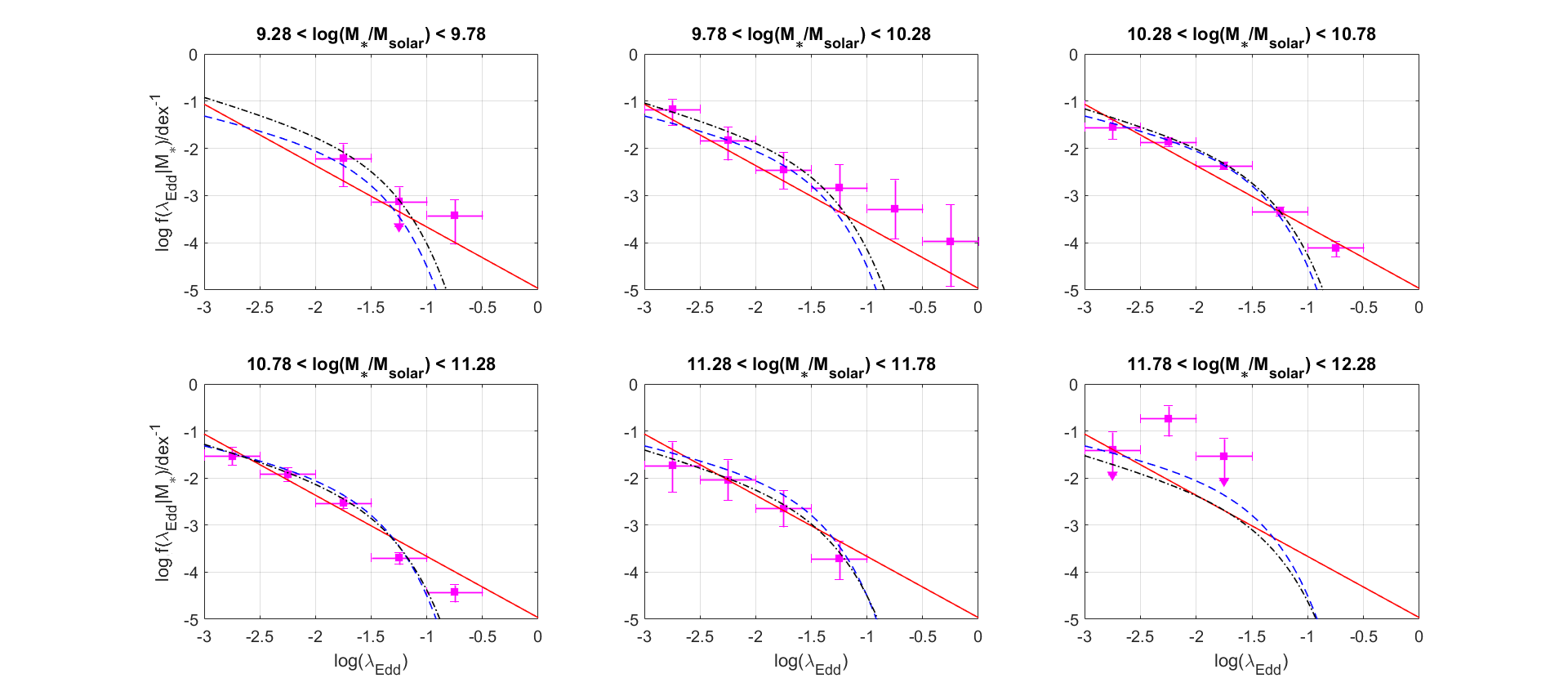}
	
	\caption{\rm \label{fitedd0} The dependence $f(\ledd)$ calculated by the $1/V_{\rm max}$ method using the rough $\Mbh$ estimates and its best fit by different models: a power law with single ($\Mstar$-independent) parameters (red solid line), a Schechter function in $\ledd$ with single
parameters (blue dashed line), and a Schechter function in $\ledd$ with a power-law dependence of the normalization on $\Mstar$ (black dotted line). }
\end{figure*}

\begin{figure*}
	\hspace{-2cm}
	\includegraphics[width=19cm]{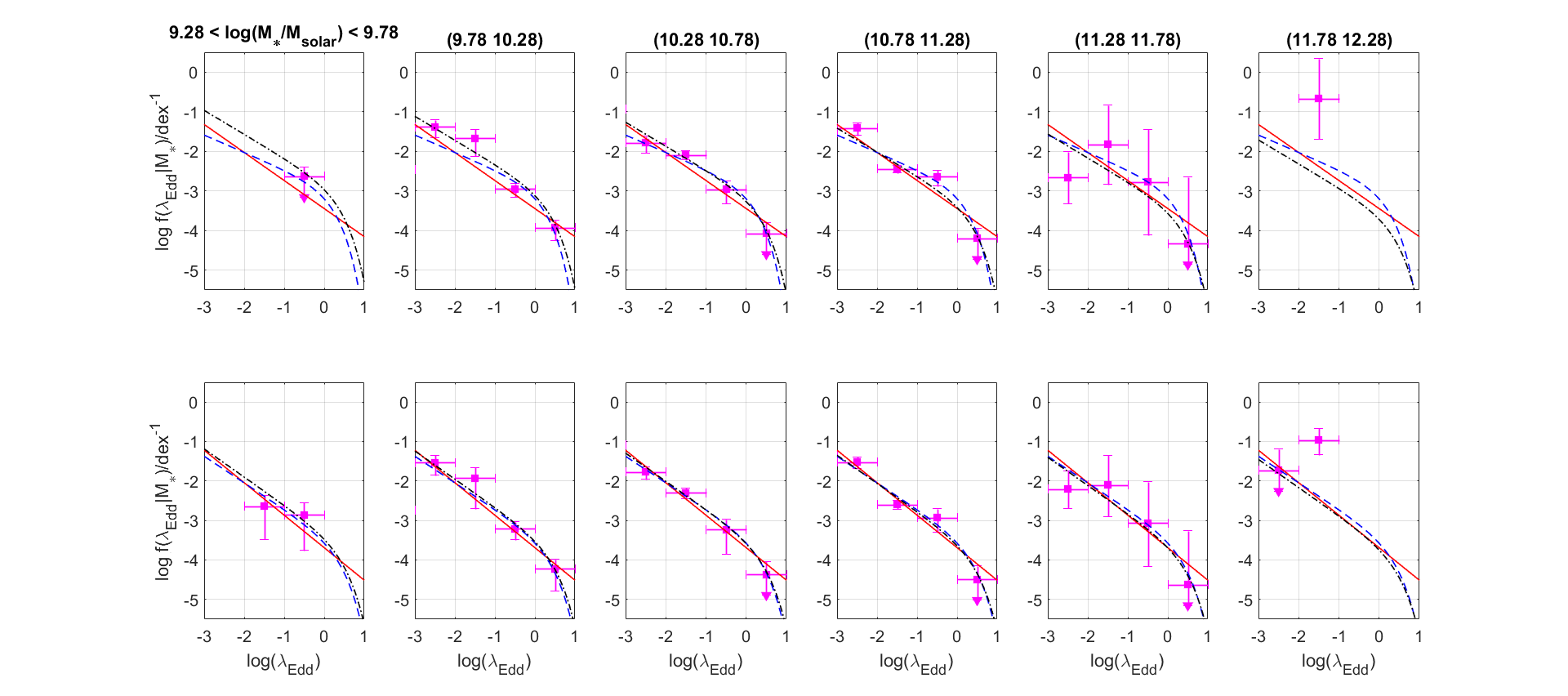}
	
	\caption{\rm \label{fitedd12} Same as Fig. \ref{fitedd0}, but when using the accurate (the upper row of graphs) and mixed (the lower row of graphs) $\Mbh$
estimates.}
\end{figure*}

Here and below, we decided to use only the range
$\lgl>-3$ because the power-law trend is expected to be smoothed out at lower Eddington ratios,
as, for example, noted in Georgakakis et al. (2014)
and Aird et al. (2018) for the earlier Universe, or
even exhibits a break. In the dependences derived
by us (see Figs. \ref{leddmbhlu} and \ref{leddmbh}) this trend is also visible
in the $\Mstar$ intervals where there are data points with
$\lgl<-3$, i.e., $10.28<\log(\Mstar/\Msun)<10.78$
and $10.78<\log(\Mstar/\Msun)<11.28$. However, based
on the available data, it is hard to say something more
specific about the pattern of the dependence $ f(\ledd)$
at $\lgl<-3$. It will be interesting to study this
question in future based on larger AGN samples.

The fitting was performed separately for three variants of the dependence $ f(\ledd)$: based on the rough,
accurate, and mixed $\Mbh$ estimates; we used $0.5\,dex$ and
$1\,dex$ $\ledd$ bins in the first case and the two remaining
ones, respectively. The results of our modeling are
shown in Figs. \ref{fitedd0} and \ref{fitedd12}, while the corresponding
parameters of the power-law dependence are given in
Table \ref{bl}.

As can be seen from the table and the graphs,
the single power-law model describes well the dependence $ f(\ledd)$ derived from the mixed $\Mbh$ estimates,
but much more poorly the dependences calculated
from the rough or accurate estimates. At the same
time, the derived parameters also vary significantly

In the next step we attempted to add an exponential cutoff at large $\ledd$ to the power-law dependence on $\ledd$, i.e., to describe the data by an analog of the Schechter function:

\begin{equation}
\label{ffitc}
f(\ledd|\Mstar)=A\left(\frac{\ledd}{\ledd^*}\right)^\gamma \exp\left(-\frac{\ledd}{\ledd^*}\right), 
\end{equation}

and then also a power-law dependence on the stellar mass:

\begin{equation}
\label{ffitksh}
f(\ledd|\Mstar)=B\left(\frac{M}{M'}\right)^\theta\left(\frac{\ledd}{\ledd^*}\right)^\gamma  \exp\left(-\frac{\ledd}{\ledd^*}\right),
\end{equation}

where $M'=10^{11}M_\odot$. The best fits by these models are shown in the same Figs. \ref{fitedd0} and \ref{fitedd12}, while the corresponding parameters are given in Tables \ref{bsh} and \ref{bksh}.

To compare the models, we used a corrected
Akaike information criterion for the $\chi^2$ distribution:

\begin{equation}
\label{aicc}
%AIC_c=2k+\chi^2+2\sum_{i} \ln{\sigma_i}+\frac{2k(k+1)}{n-k-1},
AIC_c=2k+\chi^2+\frac{2k(k+1)}{n-k-1},
\end{equation}

where $k$ is the number of model parameters, $n$ is the
number of data points, and $\chi^2$ is the $\chi^2$ value of the
model. The smaller the value of $AIC_c$, the better the
model. The derived $AIC_c$ for all of the models used
are given in the next-to-last rows of Tables \ref{bl}–\ref{bksh}.

Based on these values, we can conclude that
adding an exponential cutoff to the power-law dependence on $\ledd$ improves the quality of the fit when
using the rough $\Mbh$ estimates. For the accurate
$\Mbh$ estimates the improvement is less significant,
while for the mixed ones introducing a new parameter
does not lead to any improvement. Adding a powerlaw dependence on the galaxy stellar mass slightly
improves the quality of the fit for the rough $\Mbh$
estimates, but does not lead to any improvements for
the accurate and mixed $\Mbh$ estimates.

The Bayesian information criterion $BIC=k\ln{n}+\chi^2$ yields similar results (the lower rows in Tables \ref{bl}–\ref{bksh}), except for the insignificant improvement of the fit
for the accurate $\Mbh$ estimates when introducing a
power-law dependence on $\Mstar$.

Significant errors in the parameters $\gamma$ and $\ledd^*$ when fitting the dependences $f(\ledd)$ by a Schechter
function, especially for the model (\ref{ffitksh}), can also be
noted. The slope $\theta$ for this model applied to the mixed
$\Mbh$ estimates is consistent with zero; for the other
samples (the rough and accurate $\Mbh$ estimates) the
dependence on the stellar mass is also weak.

To summarize the results obtained, it can be noted
that adding an exponential cutoff to the power-law
dependence on $\ledd$ improves significantly the quality
of the fit only when using the rough $\Mbh$ estimates,
with the slope of the dependence (below the cutoff)
in this case being in good agreement with the slope
of $f(\ledd)$ for the accurate or mixed $\Mbh$ estimates
(for all models). Thus, the slope of $f(\ledd)$ may be
deemed to have been reliably measured in the interval
of $\lgl$ from $-3$ to $-1.5$: $\gamma=-0.7\pm0.15$. At the
same time, the position of the cutoff at $\ledd\sim 1$ (near
the critical accretion rate) cannot yet be deemed to
have been reliably found due to the existence of significant systematic uncertainties (associated mainly
with the SMBH mass estimate).

The slope of $f(\ledd)$ measured in this paper for the
local Universe is consistent, given the errors, with the
estimate obtained previously by Aird et al. (2012) for
the Universe at $0.2<z<1$: $\gamma=-0.65\pm0.04$. In addition, the normalization of the power-law dependence $f(\ledd)$ $\log{A}=-3.79\;(-3.93,\;-3.68)$\;dex
derived here also agrees with the normalization from
the mentioned paper $\log{A}_{z=0}=-3.86\pm0.1$\;dex,
given the correction for evolution based on the formula $A_z=A_{z=0.6}\,\left((1+z)/(1+0.6)\right)^{3.47\pm0.5}$, where
$\log{A}_{z=0.6}=-3.15\pm0.08$\;dex, from the same paper.

\subsection{Mean Growth Time and Duty Cycle of SMBHs at
the Present Epoch}

The derived dependence $f(\ledd)$
characterizes the distribution of current mass accretion rates onto SMBHs in the local Universe ($z<0.15$), being basically a “snapshot” of the process of
accretion onto the BHs in galactic nuclei. From this
dependence we can infer the mean SMBH growth
time at the present epoch. Indeed, although the
growth histories of different BHs could differ greatly
from one another, as a first approximation, we can
assume that the distribution of instantaneous accretion rates ($\ledd$) for a specific BH on a long time
scale ($\sim 1$--$2$ Gyr, i.e., the time interval between $z=0.15$ and $z=0$ being studied in this paper) roughly
corresponds to the present-day distribution $f(\ledd)$
for the population of SMBHs as a whole.

Let us define the characteristic BH growth time $\tau$ as

\begin{equation}
\label{tbh}
\int_0^\tau\frac{\dot{M}_{\rm BH}}{\Mbh}\,dt=1, 
\end{equation}

where

\begin{equation}
 \label{Leddt}
  \frac{\dot{M}_{\rm BH}}{\Mbh}=\frac{\Lbol}{\eta\, c^2\, \Mbh}=7.3\times10^{-16}\;\ledd
\end{equation}

(s$^{-1}$), $\eta=0.1$ is the expected accretion efficiency
onto SMBHs, and $c$ is the speed of light in a vacuum.
Here, we used relations (\ref{eq:ledd}) and (\ref{Ledd}).

By introducing the definition of the mean accretion
rate in units of the Eddington ratio,

\begin{equation}
\label{lambs}
\left<\ledd\right>\equiv\frac{1}{\tau}\int_0^\tau\ledd\,dt,
\end{equation}

we get

\begin{equation}
 \label{tdouble}
  \tau=4.4\times 10^7 \left<\ledd\right>^{-1}
\end{equation}

(years), where instead of the time averaging we
can use (in accordance with the above assumption)
the averaging over the present-day population of
SMBHs:

\begin{equation}
\label{eq:avledd}
\left<\ledd\right>=\int_{-\infty}^{+\infty}f(\ledd)\,\ledd\,d\log{\ledd}.
\end{equation}

We calculated $\tau$ from the discrete dependences
$f(\ledd)$ presented in Fig. \ref{leddmbh} in separate galaxy stellar
mass intervals using, as above, different types of
SMBH mass estimates (rough, accurate, and mixed). The quantity $\left<\ledd\right>$ was calculated by summing over
the $\ledd$ intervals, i.e.,
$\sum_{i}f(\lambda_{\rm Edd,i}|M_{j})\,\lambda_{\rm Edd,i}\,d\log{\ledd}$,
while the corresponding upper and lower limits were
calculated as
$\sum_{i}(f(\lambda_{\rm Edd,i})+\Delta f_i)\,\lambda_{\rm Edd,i}\,d\log{\ledd}$ and $\sum_{i}(f(\lambda_{\rm Edd,i})-\Delta f_i)\,\lambda_{\rm Edd,i}\,d\log{\ledd}$, where $\Delta f_i$
is the total error in the interval, including the statistical and systematic (arising when the contribution
of the active nucleus to the IR galaxy luminosity is
taken into account) ones.

\begin{figure*}
	\hspace{-2cm}
	\includegraphics[width=19cm]{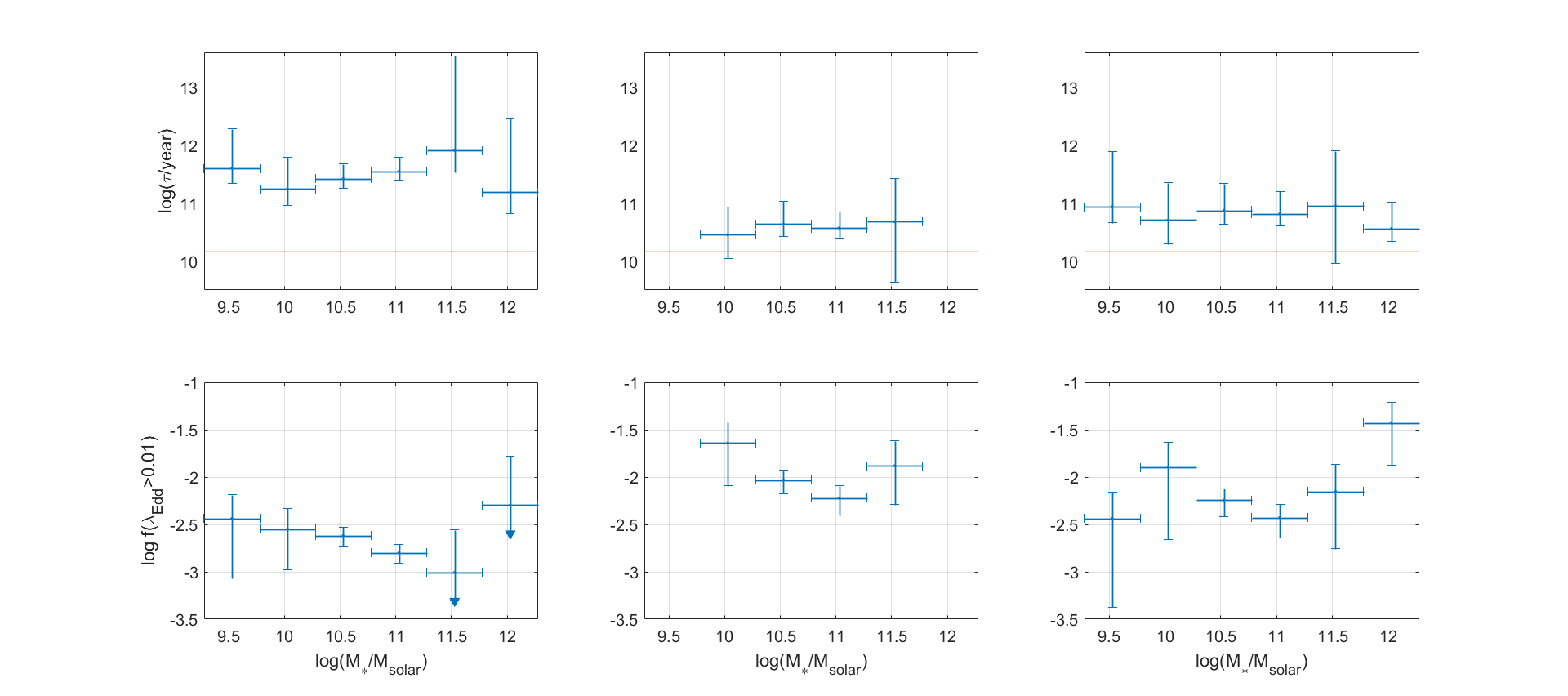}
	
	\caption{\rm \label{td} {Top:} : Characteristic SMBH growth time at the present epoch versus galaxy stellar mass when using the rough
(left), accurate (center), and mixed (right) SMBH mass estimates in the calculations by assuming the accretion efficiency
to be $\eta=0.1$ (at a different $\eta$ this time will change proportionally). The horizontal red line corresponds to the Hubble time
 $t_{\rm H}=13.7$ Gyr. {Bottom:} Duty cycle at the present epoch versus $\Mstar$ when using the rough (left), accurate (center), and mixed
(right) SMBH mass estimates in the calculations. At $\eta$ different from 0.1 the duty cycle will change inversely proportionally
to $\eta$.}
\end{figure*}

The results of our calculations are presented in
Fig. \ref{td} at the top. As we see, within the error limits, the characteristic SMBH growth time does not
depend on the galaxy stellar mass and exceeds the
lifetime of the Universe at least by several times and
may be an order of magnitude (depending on what
type of $\Mbh$ estimates is used).

Subsequently, we calculated $\tau$ based on the analytical model of $f(\ledd|\Mstar)$. Here, finding $\left<\ledd\right>$
runs into a problem when integrating over $\lgl$
in infinite limits, as required by the averaging procedure (\ref{eq:avledd}). When integrating to $+\infty$, an exponential cutoff, which lies in the interval $\lglz\in (-1.77,\;-1.58)$ for the rough estimates and $\lglz\in (0,\;1)$ for the accurate and mixed ones, solves the problem of the divergence of the power-law dependence. The existence of a cutoff at $\lglz\in (0,\;1)$  is also confirmed by works on the earlier
Universe (see, e.g., Aird et al. 2018). As $-\infty$ is
approached, the behavior of $f(\ledd|\Mstar)$ is unknown, but Fig. \ref{leddmbh} shows that this dependence flattens out or
at least has a significantly smaller slope at $\lgl<-3$. Hence it can be concluded that the value of
the integral $\int_{-\infty}^{+\infty}f_{\rm true}(\ledd|\Mstar)\,\ledd\,d\log{\ledd}$ lies
between $\int_{-3}^{+\infty}f_{\rm model}(\ledd|\Mstar)\,\ledd\,d\log{\ledd}$ and $\int_{-\infty}^{+\infty}f_{\rm model}(\ledd|\Mstar)\,\ledd\,d\log{\ledd}$, where $f_{\rm true}(\ledd|\Mstar)$
denotes the true distribution of $f(\ledd|\Mstar)$ and
$f_{\rm model}(\ledd|\Mstar)$ denotes the analytical model based
on the $\ledd$-constrained data. The uncertainty in
$f(\ledd|\Mstar)$ at small $\ledd$ affects significantly the
$\tau$ estimates based on the analytical model, but the
related error does not exceed half an order of magnitude. The uncertainties in the model parameters for
$f(\ledd|\Mstar)$ make an additional weighty contribution
to the error in the $\tau$ estimate.

If we take the analytical model (\ref{ffitc}) with the parameters from Table \ref{bsh} as a basis, then we will obtain
the following results. For the rough $\Mbh$ estimates
$\tau=3.3\;(2.7,\,4.8)\times 10^{11}$ years; for the accurate and
mixed ones $\tau=5.2\;(4.4,\,6.4)\times 10^{10}$ years and $\tau=0.9\;(0.7,\,1.1)\times 10^{11}$ years, respectively. Thus, when
using the power-law model with an exponential cutoff
at large $\ledd$, the results being obtained agree well
with those obtained above (Fig. \ref{td} at the top) by
integrating the discrete function $f(\ledd)$.

Finally, it should be noted that all of the values
of $\tau$ presented above were obtained by assuming the accretion efficiency to be 10\%, corresponding to standard accretion onto a slowly rotating BH (Shakura
and Sunyaev 1973). At a different value of the parameter $\eta$ the characteristic SMBH growth time will change proportionally.

Next, we can calculate, just as was done in Aird
et al. (2018), the SMBH “duty cycle”:

\begin{equation}
 \label{dutycycle}
  f(\ledd>0.01|\Mstar)=\int_{-2}^{+\infty}f(\ledd|\Mstar)\,d\log{\ledd},
\end{equation}

i.e., the fraction of galaxies at the present epoch in
which the SMBHs accrete matter with a rate of at
least 1\% of the critical one (assuming the accretion
to proceed via a standard disk).

We estimated $f(\ledd>0.01)$ using the discrete
dependences $f(\ledd)$ in separate $\Mstar$ intervals. The
results for different types of $\Mbh$ estimates are shown
in Fig. \ref{td}. We see that the duty cycle is $\sim 0.2$--$1$\% and
does not depend, within the error limits, on the galaxy
stellar mass.

We additionally estimated the duty cycle $f(\ledd>0.01)$ based on the analytical models (\ref{ffitc}). As a result, $f(\ledd>0.01)=0.2\;(0.18,\,0.23)\%$, $f(\ledd>0.01)=0.61\;(0.53,\,0.71)\%$, and $f(\ledd>0.01)=0.44\;(0.37,\,0.5)\%$ for the rough, accurate, and mixed $\Mbh$ estimates, respectively. These results agree well
with the estimates of $f(\ledd>0.01|\Mstar)$ based on the
discrete dependences $f(\ledd)$.

Aird et al. (2018) concluded that $f(\ledd>0.01)$
increases with $\Mstar$ for star-forming galaxies at $z\gtrsim0.5$. At the same time, no such dependence was
detected for galaxies with a low star formation rate
(at the same redshifts), except for the decrease in
$f(\ledd>0.01)$ at $11<\log(\Mstar/\Msun)<11.5$. At redshifts
$z<0.5$ the measurement errors in Aird et al. (2018)
do not allow any conclusion about the dependence
of the duty cycle on $\Mstar$ to be reached. However,
it can be seen from Fig. 6 in Aird et al. (2018) that at low $z$ for all of
the presented $\Mstar$ $f(\ledd>0.01)$ converges to $0.5$--$1\%$, in good agreement with our results. Thus,
the dependence $f(\ledd>0.01|\Mstar)$ could experience
noticeable evolution in the last several billion years.

Note that when the accretion efficiency $\eta$ differs
from 10\%, all of the above estimates of the AGN duty
cycle change inversely proportionally to $\eta$.

\section{CONCLUSIONS}

In this paper we studied the distribution of accretion rates in AGNs of the local Universe ($z<0.15$) based on homogeneous (outside the Galactic plane) near-IR (2MASS) and hard X-ray (Swift/BAT) surveys. Using sufficiently accurate SMBH mass estimates allowed us to better estimate the Eddington ratio $\ledd$ for approximately half of the AGN sample; for the remaining objects we used a less accurate estimate based on the correlation of $\Mbh$ with the galaxy stellar mass $\Mstar$. As a result, we obtained the following results for a wide range of galaxy masses, $9.28<\log(\Mstar/\Msun)<12.28$, including the most massive galaxies in the local Universe:
\begin{enumerate} 

\item 
The distribution $f(\ledd)$ above $\lgl=-3$ is described by a power law with $\Mstar$-independent parameters, declining with a
characteristic slope $\approx -0.7$ up to the Eddington limit ($\lgl\sim 0$), where there is evidence for a break.
\item
 We found evidence that at $\lgl<-3$ the dependence $f(\ledd)$ has a lower slope or flattens out.
\item
We estimated the SMBH growth time at the present epoch. It does not depend (within the error limits) on the galaxy stellar mass and exceeds the lifetime of the Universe, but by no more than an order of magnitude.
\item
We estimated the mean SMBH duty cycle (the fraction of objects with $\ledd>0.01$) at the present epoch. It does not depend (within the error limits) on $\Mstar$ either and is 0.2--1\%.
\end{enumerate} 

These results obtained for the present epoch confirm the trends found in previous studies for the earlier Universe and refine the parameters of the dependence $f(\ledd|\Mstar)$ at $z<0.15$. The revealed universal (weakly dependent on the galaxy stellar mass) pattern of the dependence $f(\ledd)$ probably stems from the fact that the main SMBH growth occurred at early epochs in the life of
the Universe, while at present the episodes of mass accretion onto SMBHs are associated mainly with stochastic processes in galactic nuclei rather than with global galaxy evolution processes.

\section*{Acknowledgements}

This study was supported by grant no. 19-12-00396 of
the Russian Science Foundation.

%%%%%%%%%%%%%%%%%%%%%%%%%%%%%%%%%

\clearpage
\thispagestyle{empty}
\onecolumn

%%%%%%%%%%%%%%%%%%%%%%%%%%%%%%%%%
\twocolumn

%%%%%%%%%%%%%%%%%%%%%%%%%%%%%%%%%
%	The end
%%%%%%%%%%%%%%%%%%%%%%%%%%%%%%%%%
\end{document}